\documentclass[preprint,aps,superscriptaddress,nofootinbib,tightenlines,floatfix]{revtex4}
\usepackage{subfigure}
\usepackage{epsfig}
\usepackage{graphicx}
\usepackage{amsmath}
\usepackage{bm}
\usepackage{verbatim}
\usepackage{color}
\usepackage{epsfig}

\def\OMIT#1{{}}

\newcommand{\gsim}{\raisebox{-0.7ex}{$\stackrel{\textstyle >}{\sim}$ }}

\newcommand{\beqa}{\begin{eqnarray}}
\newcommand{\eeqa}{\end{eqnarray}}

\newcommand{\beq}{\begin{equation}}
\newcommand{\eeq}{\end{equation}}
\newcommand{\bea}{\begin{eqnarray}}
\newcommand{\eea}{\end{eqnarray}}

\newcommand{\nn}{\nonumber \\}

\begin{document}

\begin{figure}[!t]

  \vskip -1.5cm
  \leftline{\includegraphics[width=0.25\textwidth]{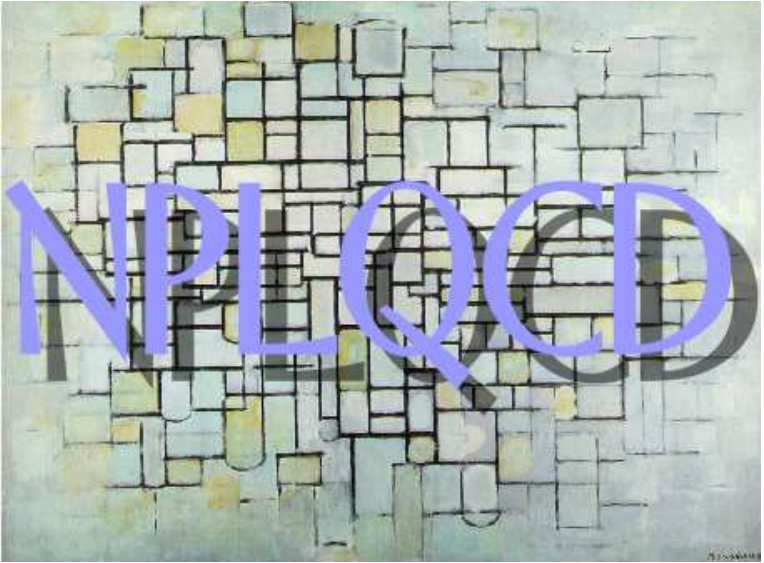}}
\end{figure}

\preprint{\vbox{ 
\hbox{UNH-11-3} 
\hbox{NT@UW-11-12}
\hbox{ICCUB-11-159}
\hbox{UCB-NPAT-11-008 } 
\hbox{NT-LBNL-11-012 } 
}}

\title{
The $I=2$ $\pi\pi$ S-wave Scattering Phase Shift from Lattice QCD}

\author{S.R.~Beane} 

\affiliation{Albert Einstein Zentrum f\"ur Fundamentale Physik,
Institut f\"ur Theoretische Physik,
Sidlerstrasse 5,
CH-3012 Bern, Switzerland}
\affiliation{Department of Physics, University
  of New Hampshire, Durham, NH 03824-3568, USA}

\author{E.~Chang}
\affiliation{Departament d'Estructura i Constituents de la Mat\`eria. 
Institut de Ci\`encies del Cosmos (ICC),
Universitat de Barcelona, Mart\'{\i} i Franqu\`es 1, E08028-Spain}

\author{W.~Detmold} 
\affiliation{Department of Physics, College of William and Mary, Williamsburg,
  VA 23187-8795, USA}
\affiliation{Jefferson Laboratory, 12000 Jefferson Avenue, 
Newport News, VA 23606, USA}

\author{H.W.~Lin}
\affiliation{Department of Physics,
  University of Washington, Box 351560, Seattle, WA 98195, USA}

\author{T.C.~Luu}
\affiliation{N Division, Lawrence Livermore National Laboratory, Livermore, CA
  94551, USA}

\author{K.~Orginos}
\affiliation{Department of Physics, College of William and Mary, Williamsburg,
  VA 23187-8795, USA}
\affiliation{Jefferson Laboratory, 12000 Jefferson Avenue, 
Newport News, VA 23606, USA}

\author{A.~Parre\~no}
\affiliation{Departament d'Estructura i Constituents de la Mat\`eria. 
Institut de Ci\`encies del Cosmos (ICC),
Universitat de Barcelona, Mart\'{\i} i Franqu\`es 1, E08028-Spain}

\author{M.J.~Savage} \affiliation{Department of Physics,
  University of Washington, Box 351560, Seattle, WA 98195, USA}

\author{A.~Torok} \affiliation{Department of Physics, Indiana University,
  Bloomington, IN 47405, USA}
\author{A.~Walker-Loud}
\affiliation{Lawrence Berkeley National Laboratory, Berkeley, CA 94720, USA}

\collaboration{ NPLQCD Collaboration }


\begin{abstract}
  \noindent 
  The $\pi^+\pi^+$ s-wave scattering phase-shift is determined below
  the inelastic threshold using Lattice QCD.  Calculations were
  performed at a pion mass of $m_\pi\sim 390~{\rm MeV}$ with an anisotropic $n_f=2+1$
  clover fermion discretization in four lattice volumes, with spatial
  extent $L\sim 2.0, 2.5, 3.0$ and $3.9~{\rm fm}$, and with a
  lattice spacing of $b_s\sim 0.123~{\rm fm}$ in the
  spatial direction and $b_t \sim b_s/3.5$ in the time direction.  The
  phase-shift is determined from the energy-eigenvalues of
  $\pi^+\pi^+$ systems with both zero and non-zero total momentum in
  the lattice volume using L\"uscher's method.  Our calculations are
  precise enough to allow for a determination of the threshold
  scattering parameters, the scattering length $a$, the effective range
  $r$, and the shape-parameter $P$, in this channel and to examine the
  prediction of two-flavor chiral perturbation theory: $m_\pi^2 a r =
  3+{\cal O}({m_\pi^2/\Lambda_\chi^2})$.  Chiral perturbation theory
  is used, with the Lattice QCD results as input, to predict
  the scattering phase-shift (and threshold parameters) at the
  physical pion mass.  Our results are consistent with
  determinations from the Roy equations and with the existing
  experimental phase shift data.
\end{abstract}
\maketitle
\tableofcontents
\vfill\eject
%

\section{Introduction}
\label{sec:intro}
\noindent

\noindent Pion-pion ($\pi\pi$) scattering at low energies is the
theoretically simplest and best-understood hadronic scattering
process.  Its simplicity and tractability follow from the
pseudo-Goldstone boson nature of the pion, a consequence of the
spontaneously broken chiral symmetry of QCD, which implies powerful
constraints on its low-momentum interactions.  The amplitudes for
$\pi\pi$ scattering are uniquely predicted at leading order (LO) in
chiral perturbation theory ($\chi$PT)~\cite{Weinberg:1966kf}.
Subleading orders in the chiral expansion 
give rise to perturbatively-small deviations from the LO
determinations (for small pion masses), and contain both calculable
non-analytic contributions and analytic terms with low-energy
constants (LEC's) that cannot be determined by chiral symmetry
alone~\cite{Gasser:1983yg,Bijnens:1995yn,Bijnens:1997vq}.
Fortunately, Lattice QCD calculations are reaching a level of
precision where statistically significant values of the LEC's in
the $I=2$ ($\pi^+\pi^+$) channel are being calculated.
Once the LEC's are obtained using unphysical lattice pion masses,
$\chi$PT can be used to predict the phase shift at the physical pion
mass to relatively high precision and with quantified uncertainties.
The current capability of Lattice QCD ---in conjuction with $\chi$PT--- to
calculate $\pi\pi$ scattering parameters very accurately is important
theoretically because
Roy-equation~\cite{Roy:1971tc,Basdevant:1973ru,Ananthanarayan:2000ht}
determinations of $\pi\pi$ scattering parameters, which use dispersion
theory to relate scattering data at high energies to the scattering
amplitude near threshold, have also reached a remarkable level of
precision~\cite{Colangelo:2001df,GarciaMartin:2011cn,Leutwyler:2006qq},
and the results of the two methods can now be compared and contrasted.

There have been independent lattice QCD determinations of the
$\pi^+\pi^+$ scattering length; with three flavors ($n_f=2+1$) of
light quarks using domain-wall valence quarks on asqtad-improved
staggered sea quarks~\cite{Beane:2005rj,Beane:2007xs}, and with two
flavors ($n_f=2$) of light quarks using twisted-mass
quarks~\cite{Feng:2009ij} and improved Wilson
quarks~\cite{Yamazaki:2003za,Yamazaki:2004qb,Aoki:2004wq,Aoki:2005uf}.
These determinations are in agreement with the Roy equation values.
The first calculation of the $\pi^+\pi^+$ scattering phase
shift was carried out by the CP-PACS collaboration, who exploited the
finite-volume strategy to study s-wave scattering with $n_f=2$
improved Wilson fermions~\cite{Yamazaki:2003za,Yamazaki:2004qb} at
pion masses in the range $m_\pi\simeq 500-1100~{\rm MeV}$.  The
amplitudes obtained from the Lattice QCD calculations were
extrapolated to the physical mass using a polynomial dependence upon
the pion mass, instead of using the known pion-mass dependence of the amplitude
based upon the symmetries of QCD encapsulated in $\chi$PT.  
In a recent paper, the Hadron Spectrum Collaboration (HSC) studied
the s-wave $\pi^+\pi^+$ phase shift with pion masses in the
range $m_\pi\simeq 390-520~{\rm MeV}$~\cite{Dudek:2010ew}.  Further, they have
provided the first Lattice QCD calculation of the $\pi^+\pi^+$ phase
shift in the d-wave ($l=2$)~\cite{Dudek:2010ew}.

In this work, which is a continuation of our high statistics Lattice
QCD
explorations~\cite{Beane:2009kya,Beane:2009gs,Beane:2009py,Beane:2011pc,Beane:2010hg},
we determine the $\pi^+\pi^+$ scattering amplitude below the inelastic
threshold.  Calculations are performed with four ensembles of
$n_f=2+1$ anisotropic clover gauge-field configurations at a single
pion mass of $m_\pi\sim 390~{\rm MeV}$ with a spatial lattice spacing
of $b_s\sim 0.123~{\rm fm}$, an anisotropy of $\xi\sim 3.5$, and with
cubic spatial volumes of extent $L\sim 2.0, 2.5, 3.0$ and $3.9~{\rm
  fm}$.  Predictions are made for a number of threshold parameters
which encode the leading momentum-dependence of the scattering
amplitude, and dictate the scattering length, effective range and
shape parameters in the effective range expansion (ERE) of the inverse
scattering amplitude.  The Lattice QCD predictions are found to be in
agreement with the Roy-equation determinations of the threshold
parameters and phase shift, and with the available experimental data.
Beyond the threshold region, the LEC's that contribute to the
two-flavor chiral expansion of the scattering amplitude are
determined, allowing for a prediction of the phase-shift at the
physical pion mass to be performed at next-to-leading order (NLO).
The predicted phase-shift is in agreement with the experimental data.

The Maiani-Testa theorem demonstrates that S-matrix elements cannot be
determined from stochastic lattice calculations of $n$-point Green's functions at
infinite volume, except at kinematic thresholds~\cite{Maiani:1990ca}.
L\"uscher showed that by
computing the energy levels of two-particle states in the
finite-volume lattice, the $2\rightarrow 2$ scattering amplitude can
be
recovered~\cite{Huang:1957im,Hamber:1983vu,Luscher:1986pf,Luscher:1990ux,Beane:2003da,Rummukainen:1995vs,Kim:2005gf,Christ:2005gi,Feng:2011ah,Lang:2011mn}.
These energy levels are found to deviate from those of two
non-interacting particles by an amount that depends on the scattering
amplitude (evaluated at that energy) and varies inversely with the
lattice spatial volume in asymptotically large volumes.
In this paper, L\"uscher's method is used to extract the phase shift
from the lattice-determined energy-levels.

This paper is organized as follows. In Sec.~\ref{sec:latticedet},
we provide some details of the lattice calculations: we discuss the
anisotropic clover lattices that are used and the determination of the
anisotropy parameter.  Sec.~\ref{sec:Lumethod} gives a summary of
the eigenvalue equation which is relevant to extracting phase shifts
from lattice-measured energy levels, in the center-of-mass (CoM) system and
in boosted (lattice = ``laboratory'') systems.  The results of the
Lattice QCD calculations are presented in
Sec.~\ref{sec:pipiscattering} and relevant fits that are used to
determine the effective range parameters, up to and including the
shape parameter, are discussed.  Sec.~\ref{sec:chiextrap} includes
a summary of the relevant $\chi$PT formulas, the chiral fits to the
lattice data, and the prediction for the $\pi^+\pi^+$ phase shift up
to the inelastic threshold at the physical pion mass.  Finally, a
summary of our predictions and a discussion of the
systematic uncertainties is given in Sec.~\ref{sec:disc}.

\section{Details of the Lattice QCD Calculations}
\label{sec:latticedet}
\noindent

\subsection{Anisotropic Clover Lattices}

\noindent Anisotropic gauge-field configurations have proven useful
for the study of hadronic
spectroscopy~\cite{Dudek:2009qf,Bulava:2009jb,Lin:2008pr,Edwards:2008ja},
and, as the calculations required for studying multi-hadron systems
rely heavily on spectroscopy, we have put considerable effort into
calculations using ensembles of gauge fields with clover-improved
Wilson fermion actions with anisotropic lattice spacing that have been
generated by the HSC.  In particular, the $n_f=2+1$ flavor anisotropic
clover Wilson action~\cite{Okamoto:2001jb,Chen:2000ej} with stout-link
smearing~\cite{Morningstar:2003gk} of the spatial gauge fields in the
fermion action with a smearing weight of $\rho=0.14$ and $n_\rho=2$
has been used.  The gauge fields entering the fermion action are not
smeared in the time direction, thus preserving the ultra-locality of
the action in the time direction.  Further, a tree-level
tadpole-improved Symanzik gauge action without a $1\times 2$ rectangle
in the time direction is used.

The present calculations are performed on four ensembles of
gauge-field configurations with $L^3\times T$ of $16^3\times 128$,
$20^3\times 128$, $24^3\times 128$ and $32^3\times 256$ lattice sites,
with a renormalized anisotropy $\xi=b_s/b_t$ where $b_s$ and $b_t$ are
the spatial and temporal lattice spacings, respectively. The spatial
lattice spacing of each ensemble is $b_s = 0.1227\pm 0.0008~{\rm
  fm}$~\cite{Lin:2008pr} giving spatial lattice extents of $L\sim 2.0,
2.5, 3.0$ and $3.9~{\rm fm}$ respectively.  The same input light-quark
mass parameters, $b_t m_l = -0.0840$ and $b_t m_s = -0.0743$, are used
in the production of each ensemble, giving a pion mass of $m_\pi\sim
390~{\rm MeV}$.  The relevant quantities to assign to each ensemble
that determine the impact of the finite lattice volume and temporal
extent are $m_\pi L$ and $m_\pi T$, which are given in
Table~\ref{tab:LQCDpionmasses}.  In addition, we tabulate the pion
masses on the four lattice volumes. As discussed in detail in
Ref.~\cite{Beane:2011pc}, exponential finite-volume corrections to the
pion masses are negligible for these volumes, a necessary condition
for the application of L\"uscher's finite-volume method for obtaining
phase shifts. Additionally, the predicted exponential finite-volume
corrections to $\pi\pi$ scattering near threshold are expected to be
negligible~\cite{Bedaque:2006yi}.  Multiple light-quark propagators
were calculated on each configuration in the four ensembles. The
source locations were chosen randomly in an effort to minimize
correlations among propagators.

\begin{table}[!ht]
  \caption{Results from the Lattice QCD calculations in the 
     four lattice volumes.  ${\rm t.l.u}$ denotes temporal lattice units.
  }
  \label{tab:LQCDpionmasses}
  \begin{ruledtabular}
    \begin{tabular}{c||cccc}
      $L^3\times T$  &  $16^3\times 128$ &  $20^3\times 128$ &  $24^3\times 128$ &
      $32^3\times 256$  \\
      \hline
      $L~({\rm fm})$ & $\sim$2.0 &  $\sim$2.5 &  $\sim$3.0 &  $\sim$3.9 \\
      $m_\pi L$ & 3.888(20)(01) & 4.8552(84)(35) & 5.799(16)(04) & 7.7347(74)(91)\\
      $m_\pi T$ & 8.89(16)(01) & 8.878(54)(22) & 8.836(85)(02) & 17.679(59)(73)\\
      $m_\pi$ (t.l.u.) &  0.06943(36)(0) & 0.06936(12)(0) & 0.06903(19)(0) & 0.069060(66)(81)\\
    \end{tabular}
  \end{ruledtabular}
\end{table}

\subsection{Determination of the Anisotropy Parameter, $\xi$}
\noindent 
In the continuum and in infinite-volume, the energy-momentum relation
for the pion is that of special relativity, $E^2=m_\pi^2 + |{\bf
  p}|^2$.  In Lattice QCD calculations, this relation is
more complicated due to the finite lattice spacing
(including the violation of Lorentz invariance) and the finite-volume,
resulting in $E^2$ being a non-trivial function of ${\bf p}$, which has
a polynomial expansion at small momentum. Retaining the leading terms
in the energy-momentum relation, including the lattice anisotropy
$\xi$, the energy and mass in temporal lattice units, and the momentum
in spatial lattice units (${\rm s.l.u}$) are related by
\begin{eqnarray}
\left(\,b_t\,E_\pi\left(|{\bf n}|\right)\right)^2 
& = &
(b_t\,m_\pi)^2 \ +\
\frac{1}{\xi^2}\,\left(\frac{2\,\pi\,b_s}{L}\right)^2\,{\bf n}^2 
\ .
\end{eqnarray}
The Lattice QCD calculations of the energy of the single pion state at
a given momentum ${\bf p}=\frac{2\pi}{L}{\bf n}$ (where ${\bf n}$ is
an integer triplet) allows for a determination of $\xi$, and hence
establishes the single-particle energy-momentum relation that is crucial for
determining the scattering amplitude from the location of two-particle
energy eigenvalues.  We obtain $\xi=3.469(11)$ where the statistical
and systematic uncertainties have been combined in quadrature.  This
is consistent with the value determined by Dudek {\it et al}. of
$\xi=3.459(4)$~\cite{Dudek:2010ew}.  A fit to a higher order
polynomial provides a result that is consistent with this value but
with larger uncertainties in the contributing terms.  It is important
to use the lattice determined value of $\xi$, and to propagate its
associated uncertainty, as small variations in this parameter are
amplified in the determination of the scattering amplitude from
two-particle energy-eigenvalues when the interaction is weak (and the
energy of the two-particle state is consequently near that of the non-interacting
system).

\section{The Finite Volume Methodology}
\label{sec:Lumethod}
\noindent

\noindent 
The formalism that was put in place by L\"uscher to extract
two-particle scattering amplitudes below the inelastic threshold from
the energy-eigenvalues of two-particle systems at rest in a finite
cubic volume~\cite{Luscher:1986pf,Luscher:1990ux} was extended to
systems with non-zero total momentum by Rummukainen and
Gottlieb~\cite{Rummukainen:1995vs}.  Subsequent derivations have
verified and extended~\cite{Yamazaki:2004qb, Kim:2005gf,Christ:2005gi,Feng:2011ah,Lang:2011mn}
the work in that paper.  The use of boosted systems allows for the
amplitude to be determined at more values of momentum (in the CoM),
between those defined by ${2\pi\over L} {\bf n}$.  Here the
results that are relevant to the present analysis of the boosted
$\pi^+\pi^+$ systems, and to systems at rest, are restated.

Using the notation of Ref.~\cite{Kim:2005gf}, the energy in the
CoM frame is denoted by $E^*$, which is related to
the energy $E$ and momentum ${\bf P}_{cm}$ in the ``laboratory
system'' (the total lattice momentum) by $E^{* 2} = E^2 - |{\bf
  P}_{cm}|^2$.  In what follows, it is useful to define $P_{cm} =
|{\bf P}_{cm}|$.  The $\gamma$-factor is straightforwardly defined by
$\gamma = E/E^*$, and $E^*$ is also related to the magnitude of the
momentum of each $\pi^+$ in the CoM frame $q^*$ by $E^{* 2} = 4\left[\
  q^{* 2} + m_\pi^2\ \right]$.  The real part of the inverse of the
s-wave scattering amplitude below inelastic threshold, and hence the
scattering phase-shift, can be extracted from the total energy of the
two-particle system with total momentum ${\bf P}_{cm} = {2\pi\over
  L}{\bf d}$ in the finite-volume via the generalized L\"uscher
eigenvalue relation
\begin{eqnarray}
q^*\cot\delta(q^*)
& = & 
{2\over \gamma L \sqrt{\pi}}\ 
Z_{00}^{\bf d}(1; \tilde q^{*2})
\ \ \ ,
\label{eqn:RumGott}
\end{eqnarray}
where the dimensionless quantity $\tilde q^{*}$ 
is defined by $\tilde q^{*} = {L\over 2\pi} q^*$.
The function $Z_{00}^{\bf d}(1; \tilde q^{*2})$ is a generalization of the
functions defined by L\"uscher~\cite{Luscher:1986pf,Luscher:1990ux},
\begin{eqnarray}
Z_{LM}^{\bf d}(1; \tilde q^{*2}) &  = &   
\sum_{\bf r}
{|{\bf r}|^L\ Y_{LM}(\Omega_{\bf r})
\over |{\bf r}|^2-\tilde q^{* 2} }\ 
\ \ \ ,
\label{eqn:ZLMdef}
\end{eqnarray}
where the $Y_{LM}$ are spherical harmonics and the sum is over vectors defined by 
\begin{eqnarray}
{\bf r}
& = & 
{1\over\gamma}\left( {\bf n}_\parallel - {1\over 2} {\bf d} \right)
+ {\bf n}_\perp
\ =\ 
\hat\gamma^{-1}({\bf n} - {1\over 2}{\bf d})
\ \ \ ,
\label{eq:rdef}
\end{eqnarray}
which in turn are related to the lattice momentum-vectors by ${\bf k}
= {2\pi\over L}{\bf n} = {2\pi\over L}\left( {\bf n}_\parallel + {\bf
    n}_\perp \right)$.  The ${\bf n}$ are triplets of integers and the
decomposition of ${\bf n}$ is along the direction defined by the
boost-vector ${\bf d}$.  L\"uscher presented a
method~\cite{Luscher:1986pf,Luscher:1990ux} which can be
used~\cite{Rummukainen:1995vs} to accelerate the numerical evaluation
of the sum in eq.~(\ref{eqn:ZLMdef}), and a generalization of that method
leads to
\begin{eqnarray}
Z_{LM}^{\bf d}(1; \tilde q^{*2}) &  = &   
\sum_{\bf r}
{e^{-\Lambda (|{\bf r}|^2-\tilde q^{*2})}
\over |{\bf r}|^2-\tilde q^{*2}}\ 
|{\bf r}|^L\ Y_{LM}(\Omega_{\bf r})
\nonumber\\
& 
\ +\ 
&
\delta_{L,0}\ Y_{00}\ 
\gamma \pi^{3/2}\ 
\left[\ 
2 \tilde q^{*2}\int_0^\Lambda \ dt\  
{e^{t \tilde q^{*2}}\over \sqrt{t}}
\ -\ {2\over\sqrt{\Lambda}} e^{\Lambda \tilde q^{*2}}
\right]
\nonumber\\
&
\ +\
& 
\gamma 
\sum_{{\bf w}\ne {\bf 0}}\ 
e^{-i\pi {\bf w}\cdot {\bf d}}\ 
|\hat\gamma{\bf w}|^L\ 
 Y_{LM}(\Omega_{\hat\gamma{\bf w}})\ 
\int_0^\Lambda\ dt\ 
\left({\pi\over t}\right)^{3/2+L}\ e^{t \tilde q^{*2}}\ 
e^{-{\pi^2 |\hat\gamma{\bf w}|^2\over t}}
\ \ \ ,
\label{eq:ZLM}
\end{eqnarray}
where
\begin{eqnarray}
\hat\gamma{\bf w}
& = & 
\gamma {\bf w}_\parallel + {\bf w}_\perp
\ \ \ .
\end{eqnarray}
The value of the sum is independent of the choice of $\Lambda$, and 
$\Lambda=1$ has been used in previous works~\cite{Yamazaki:2004qb}.

The energy-level structure resulting from $Z_{00}^{\bf d}(1; \tilde
q^{*2})$ has been discussed previously,
e.g. Ref.~\cite{Rummukainen:1995vs}.  For the present calculations of
boosted systems it is important to identify the closely spaced
energy-levels.  This is because the amplitudes extracted from such
levels are subject to large systematic and statistical uncertainties
due to the rapid variation of $Z_{00}^{\bf d}(1; \tilde q^{*2})$ in
their vicinity, and also due to the difficulty in separating the
states contributing to the correlation functions.  The energy-levels
associated with two non-interacting particles are located at the poles
of $Z_{00}^{\bf d}(1; \tilde q^{*2})$, and eq.~(\ref{eq:rdef}) gives
\begin{eqnarray}
{\bf d} = (0,0,0) & : & {\tilde q^{*2}} \ =\ 0, \ 1, \ 2, \ 3, \ 4, \ 5, 
\ ....
\nonumber\\
{\bf d} = (0,0,1) & : & {\tilde q^{*2}} \ =\ {1\over 4\gamma^2}, \ 
{4\gamma^2+1\over 4\gamma^2}, 
\ 
\underbrace{{  {9\over 4\gamma^2}  }, \ { {8\gamma^2+1\over 4\gamma^2}  }} , \ {4\gamma^2+9\over 4\gamma^2}, \ \ {16\gamma^2+1\over 4\gamma^2}, 
\ ....
\nonumber\\
{\bf d} = (0,1,1) & : & {\tilde q^{*2}} \ =
\ \underbrace{{ {1\over 2\gamma^2} }, 
\ { {1\over 2} }},
\ \underbrace{{  {2\gamma^2+1\over 2\gamma^2} }, 
\ { {3\over 2} }}, 
\ \underbrace{{ {4\gamma^2+1\over 2\gamma^2} },
\ { {4+\gamma^2\over 2\gamma^2}  }}
\ ....
\ \ \ ,
\label{eq:nonintlevs}
\end{eqnarray}
and so forth, where the underbraces denote states that become
degenerate as $\gamma\rightarrow 1$. We stress that the relations
summarized in this section are only valid below inelastic threshold.

\section{$\pi^+\pi^+$ Scattering on the Lattice}
\label{sec:pipiscattering}
\subsection{Lattice Phase Shift}
\
\begin{figure}[!ht]
  \centering
     \includegraphics[width=0.65\textwidth]{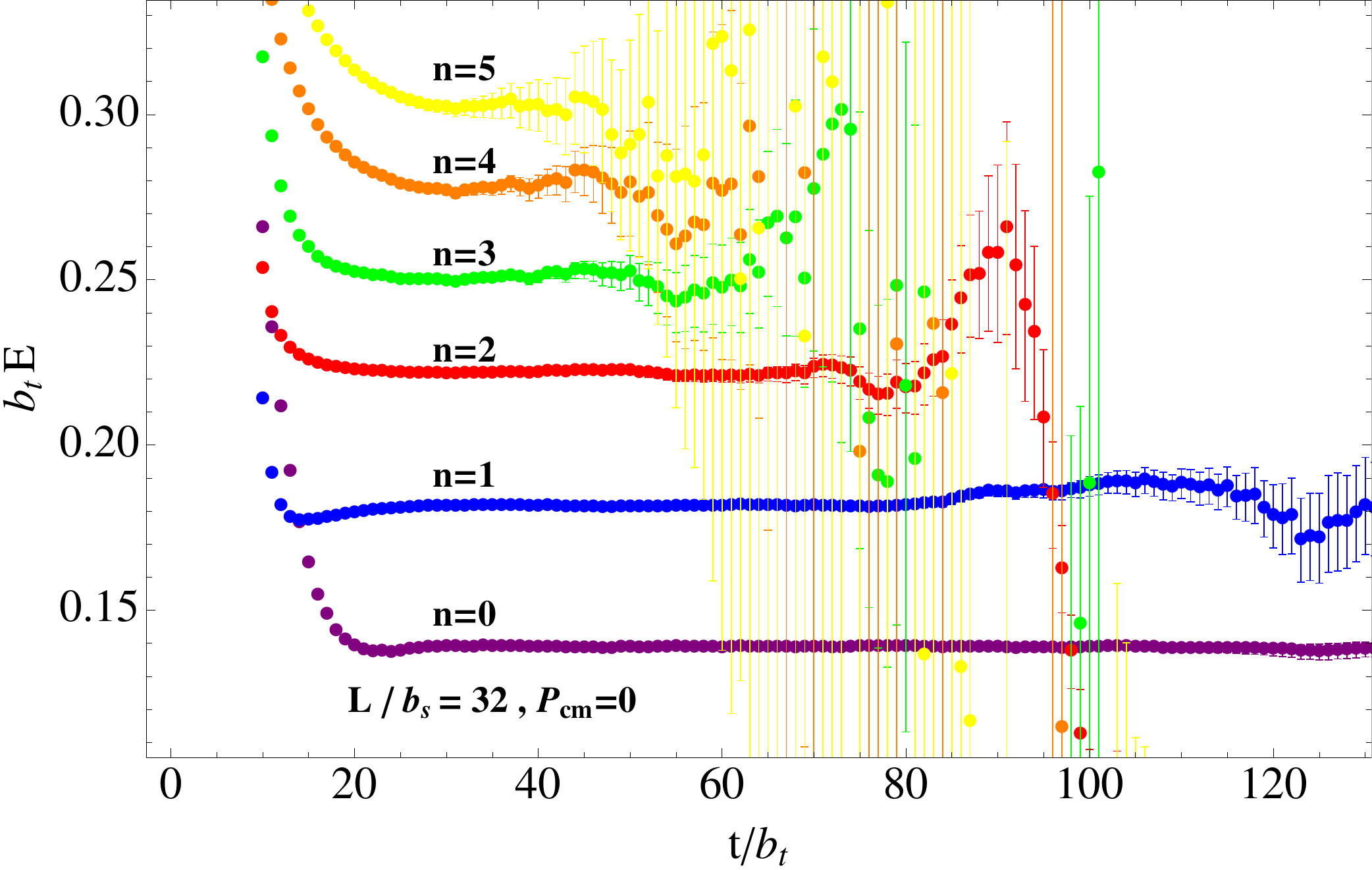} 
     \includegraphics[width=0.65\textwidth]{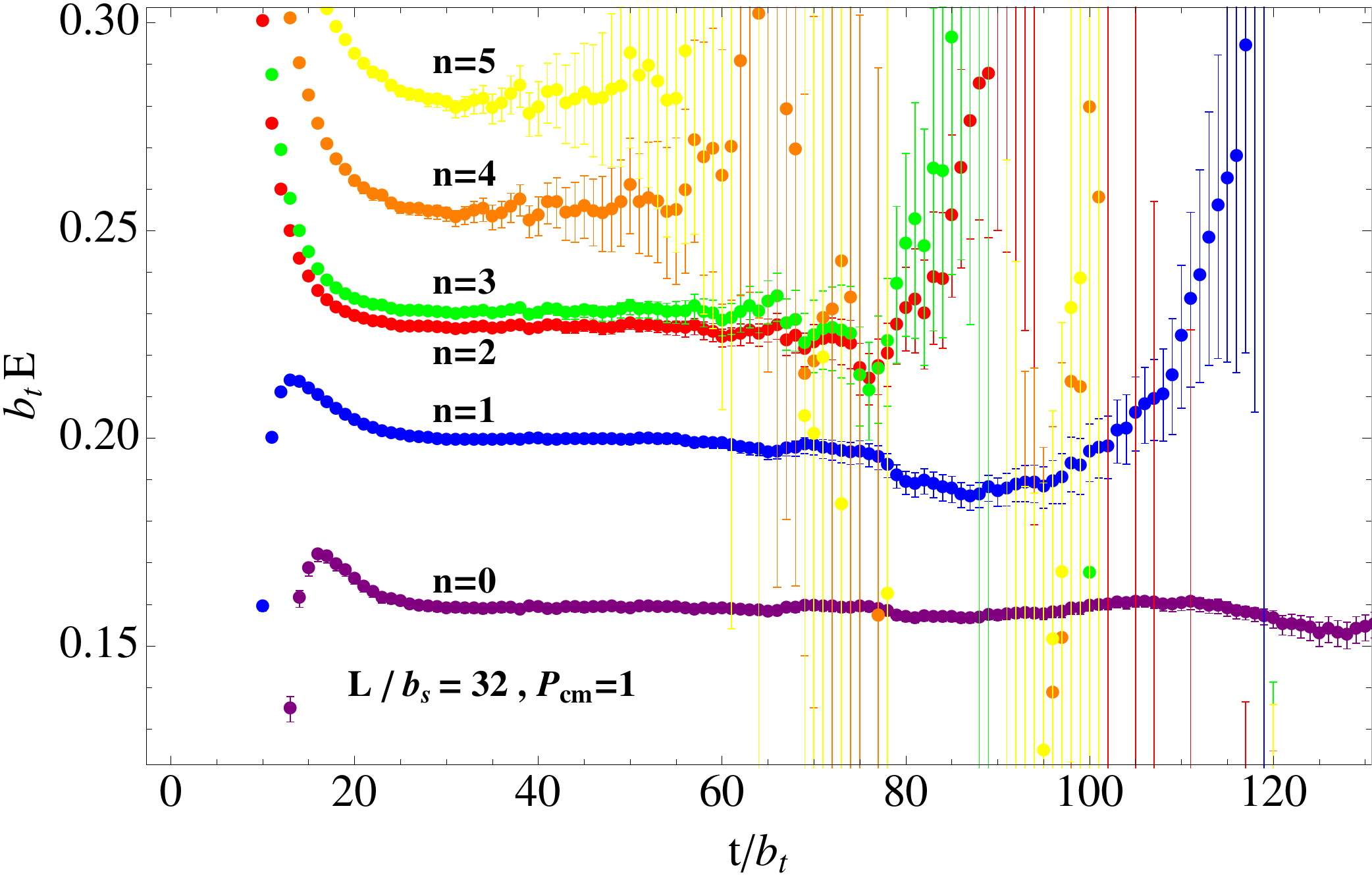} 
     \includegraphics[width=0.65\textwidth]{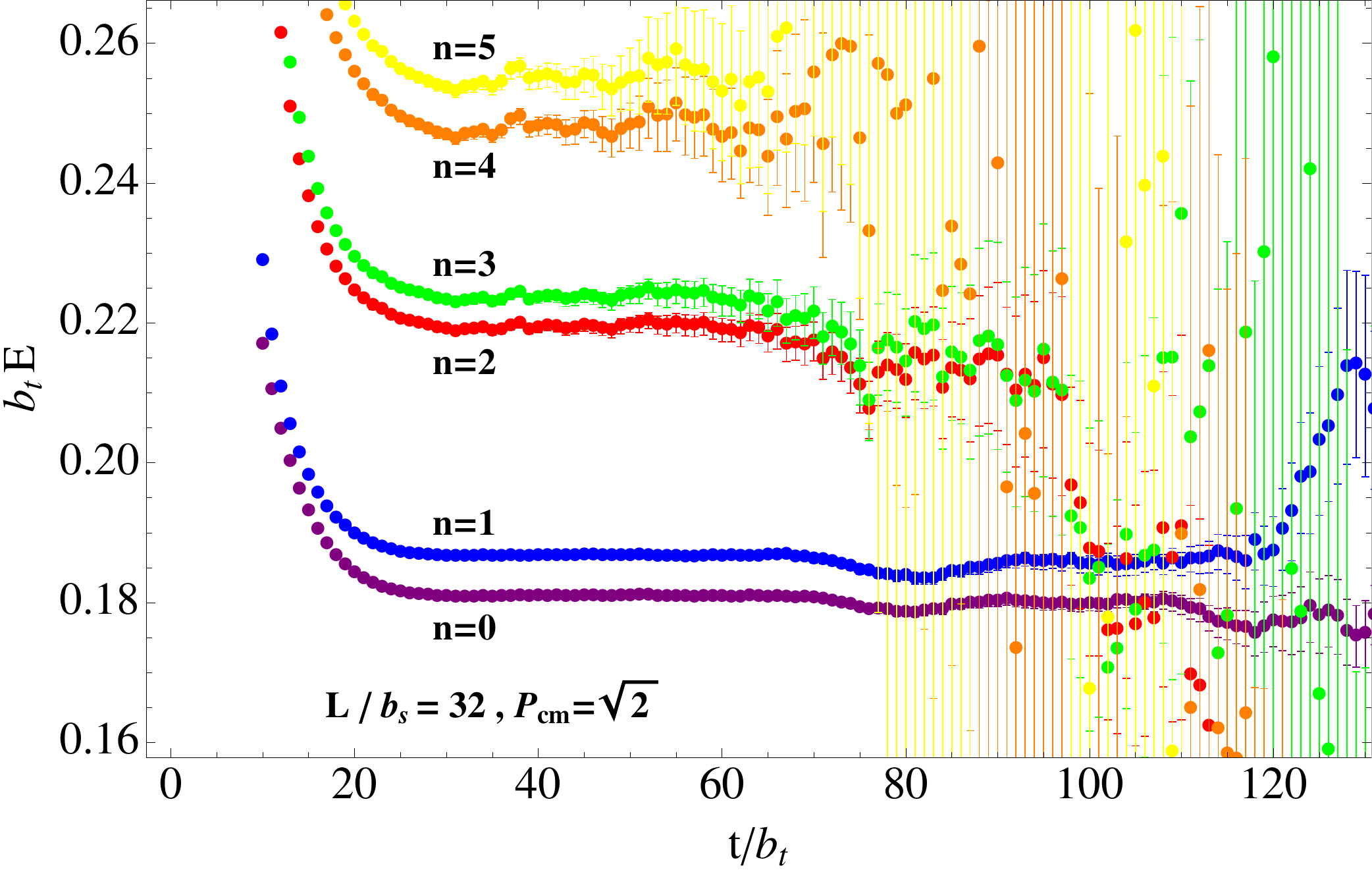}
     \caption{The two-pion EMP's for the first six levels (here $n$ indicates the level) with
       $P_{cm}=0$ (top), $1$ (middle) and $\sqrt{2}$ (bottom)
       in units of the temporal lattice spacing on the
       $32^3\times 256$ ensemble. Only one half of the temporal
       lattice points are shown. }
\label{fig:elevels32}
\end{figure}

\begin{figure}[!ht]
  \centering
     \includegraphics[width=0.84\textwidth]{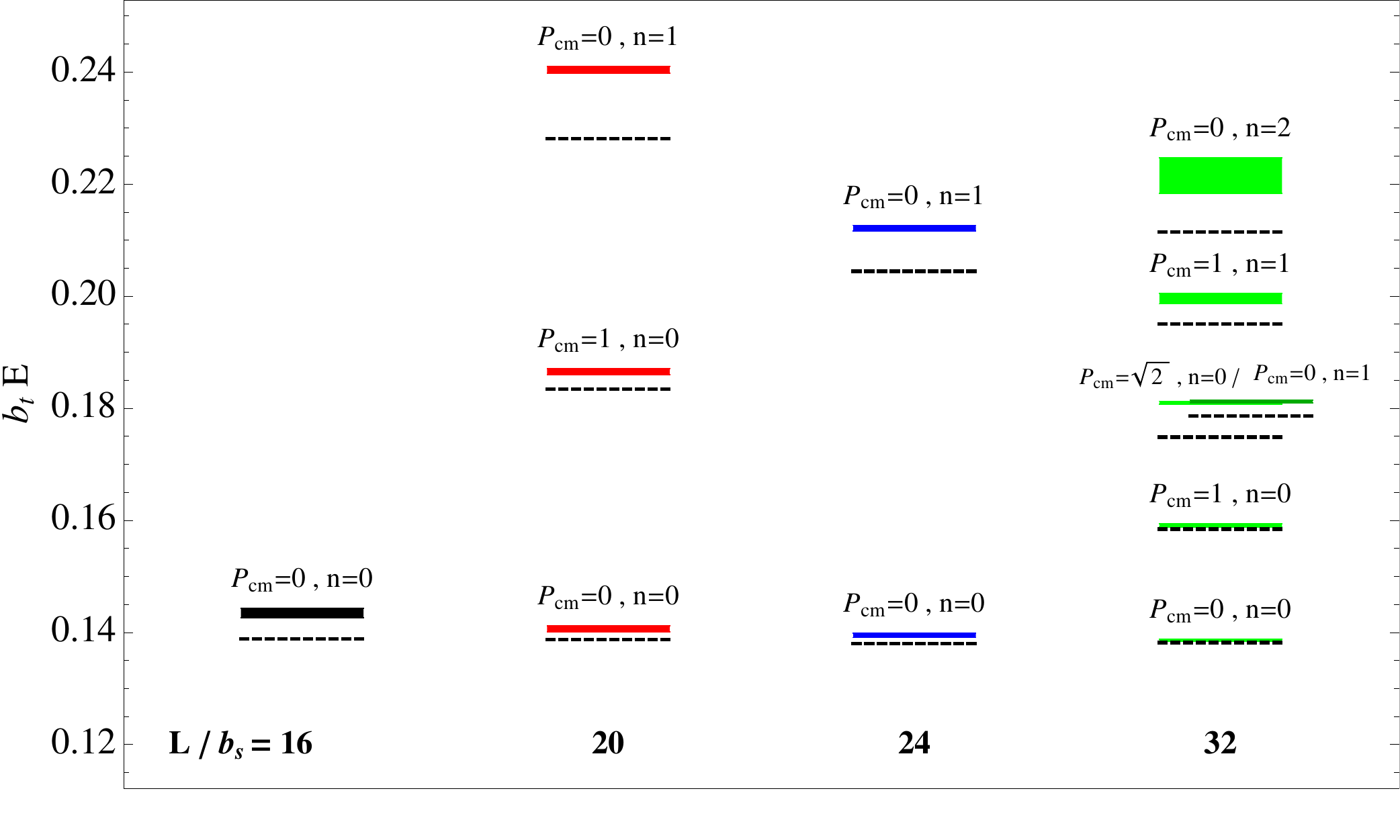}\\
\caption{The two-pion energies in units of the temporal lattice spacing for the
  lattice ensembles
considered in this work. The (vertical) thickness of each level indicates the
uncertainty of the energy determination.
Each state is labeled according to its center-of-mass momentum $P_{cm}$, and its excitation
level $n$. The non-interacting levels are denoted by dashed (black) lines. 
Notice that the $32^3\times 256$ $P_{cm}=\sqrt{2},n=0$ and $P_{cm}=0,n=1$ levels are nearly
degenerate.}
\label{fig:elevels}
\end{figure}

\begin{figure}[!ht]
  \centering
     \includegraphics[width=0.93\textwidth]{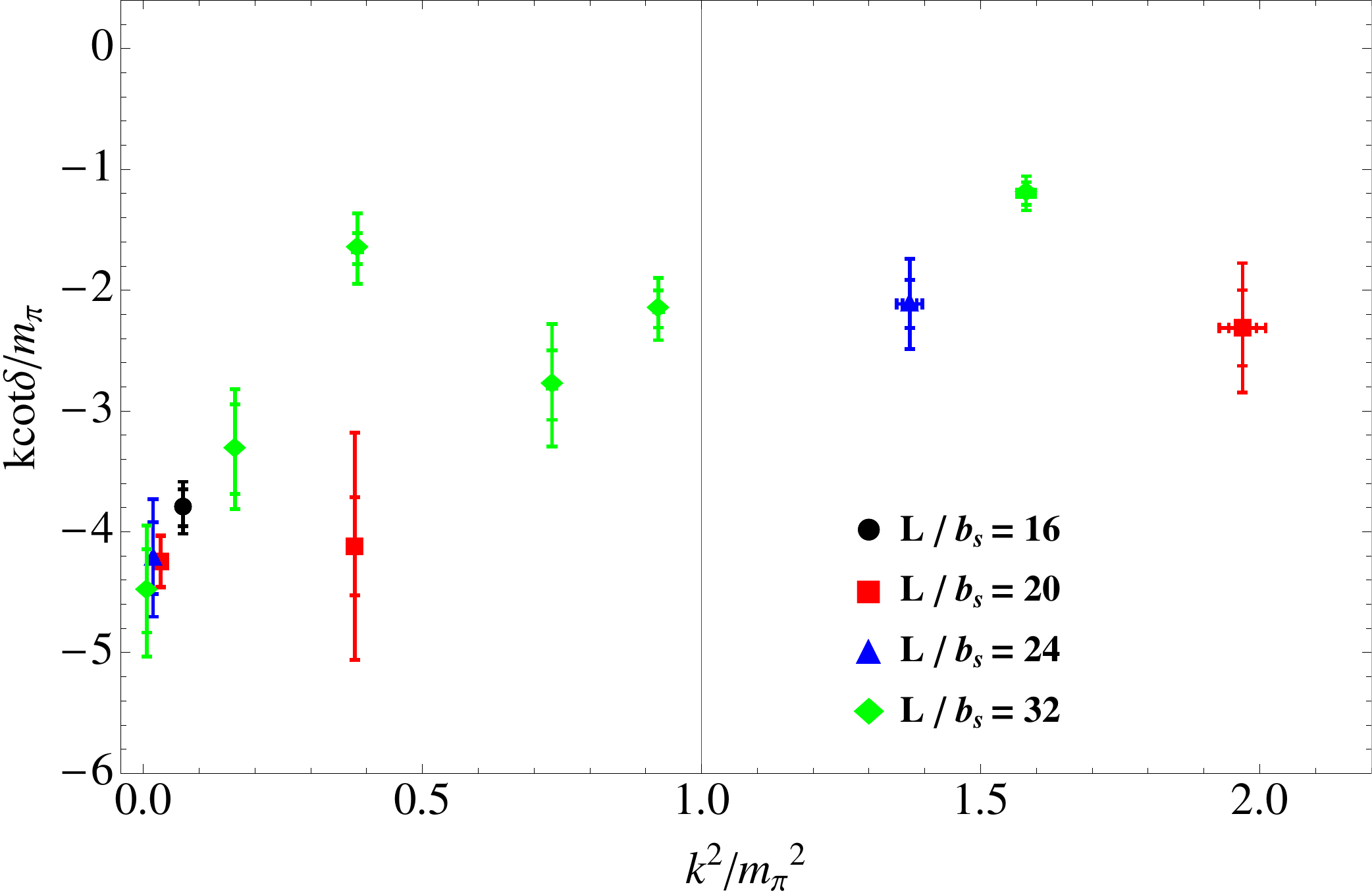}
\caption{Results of the Lattice QCD calculations
processed through the energy-eigenvalue relation to give values of 
$k\cot\delta/m_\pi$. The \{circles, squares, triangles, diamonds\}
(\{black, red, blue, green\}) correspond to 
the ensembles 
\{$16^3\times 128$, $20^3\times 128$, $24^3\times 128$, $32^3\times 256$\}. 
Statistical and systematic uncertainties are shown as the inner and outer
error-bars, respectively.
The vertical (blue) line at $k^2=m_\pi^2$ indicates the limit of the 
range of validity of the ERE  set by the t-channel cut.
The inelastic threshold is at $k^2=3 m_\pi^2$.
}
\label{fig:kcotdeltadata}
\end{figure}

\noindent 
The scattering of pions in the $I=2$ channel is perturbative at low
momentum and at small light-quark masses, as guaranteed by
$\chi$PT. In a finite volume, this translates into two-pion energies
that deviate only slightly from the non-interacting energies; i.e.,
the sum of the pion masses (or boosted pion masses for moving
systems).  We have analyzed $\pi^+\pi^+$ correlation functions with
$P_{cm}=0,1,\sqrt{2}$ and with various (non-interacting) momentum
projections among the pions.  It is straightforward to partially
diagonalize this system of correlation functions into the
energy-eigenstates at intermediate and long times. This is achieved by
assuming that the two-pion energy levels are close to their
non-interacting values, and then varying the linear combination of
correlation functions in order to maximize the plateau
region. (Coupled exponential fits to the various correlators with the
same $P_{cm}$ lead to consistent determinations.)  As an example, in
fig.~\ref{fig:elevels32} we show the two-pion effective mass plots
(EMP's) on the $32^3\times 256$ ensemble with $P_{cm}=0,1,\sqrt{2}$.
Six energy levels can be clearly identified in the EMP's in
fig.~\ref{fig:elevels32} for each of the values of $P_{cm}$. (Note
that these levels clearly show the near degeneracies of the
non-interacting system as established in eq.~(\ref{eq:nonintlevs}).)
However, only the first few levels, when propagated through the
eigenvalue equation, lead to statistically significant values for the
phase shift. While the energies of other levels are established, the
structure of the eigenvalue equation is such that the uncertainties,
as small as they appear, are sufficiently large to produce
uncertainties in the amplitude that are too large and preclude
statistical significance\footnote{The EMPs of fig~\ref{fig:elevels32}
  indicate that the signal-to-noise ratio of the two-pion correlation
  functions decreases with increasing excitation number.}.  The states that have been
analyzed to produce amplitudes and phase-shifts are given in Table~\ref{tab:LQCDpionpion}, and are shown
in fig.~\ref{fig:elevels}.  Note that momenta are quoted in units of
$m_\pi$ in order to formulate the subsequent analysis in a manner that
is independent of the scale setting.
\begin{table}[!ht]
  \caption{Results from the Lattice QCD calculations 
    of $\pi^+\pi^+$ scattering in the four lattice volumes. 
$P_{\rm cm}$ denotes the magnitude of the momentum of the center-of-mass in
    units of $2\pi/L$.
In the column denoted by ``level'', 
{\rm g.s.} denotes the ground state, 1st denotes the first excited state and
2nd denotes the second excited state.
  }
  \label{tab:LQCDpionpion}
  \begin{ruledtabular}
    \begin{tabular}{cccccc}
$k^2/m_\pi^2$  & $L^3\times T$  & ${P}_{\rm cm}$ & {\rm level} & $k\cot\delta/m_\pi$ & $\delta $
\\
      \hline
0.00678(54)(81)  & $32^3\times 256$ & 0 & $n=0$& -4.49(35)(52) & -1.06(12)(18)
\\
0.01772(14)(23)  & $24^3\times 128$ & 0 & $n=0$& -4.24(32)(49) & -1.82(19)(30)
\\
0.0309(17)(27)   & $20^3\times 128$ & 0 & $n=0$& -4.25(21)(34) & -2.37(18)(29)
\\
0.0715(32)(48)   & $16^3\times 128$ & 0 & $n=0$& -3.80(15)(22) & -4.03(25)(35)
\\
0.1641(20)(23)   & $32^3\times 256$ & 1 & $n=0$& -3.33(38)(48) & -7.1(0.8)(1.0)
\\
0.378(5)(11)  & $20^3\times 128$ & 1  & $n=0$& -4.1(0.4)(1.0) & -8.6(0.8)(3.6)
\\
0.3838(42)(85)   & $32^3\times 256$ & $\sqrt{2}$ & $n=0$& -1.65(12)(28) & -20.6(1.5)(3.1)
\\
0.7323(53)(88)   & $32^3\times 256$ & 0 & $n=1$ & -2.78(29)(57) & -17.2(1.7)(2.9)
\\
0.9233(51)(73)   & $32^3\times 256$ & 1 & $n=1$ & -2.14(16)(26) & -24.1(1.6)(2.6)
\\
1.373(13)(22)   & $24^3\times 128$ & 0 & $n=1$& -2.10(19)(36) & -29.2(2.3)(4.3)
\\
1.582(9)(16)   & $32^3\times 256$ & 0 & $n=2$& -1.19(09)(14) & -46.5(2.3)(3.5)
\\
1.969(02)(04)    & $20^3\times 128$ & 0 & $n=1$ & -2.33(32)(56)  & -31.6(3.5)(5.6)
\\
    \end{tabular}
  \end{ruledtabular}
\end{table}
The values of $k\cot\delta/m_\pi$ and $\delta$ resulting from the energy-eigenvalues
are shown in fig.~\ref{fig:kcotdeltadata}. 
\begin{figure}[!ht]
  \centering
     \includegraphics[width=0.93\textwidth]{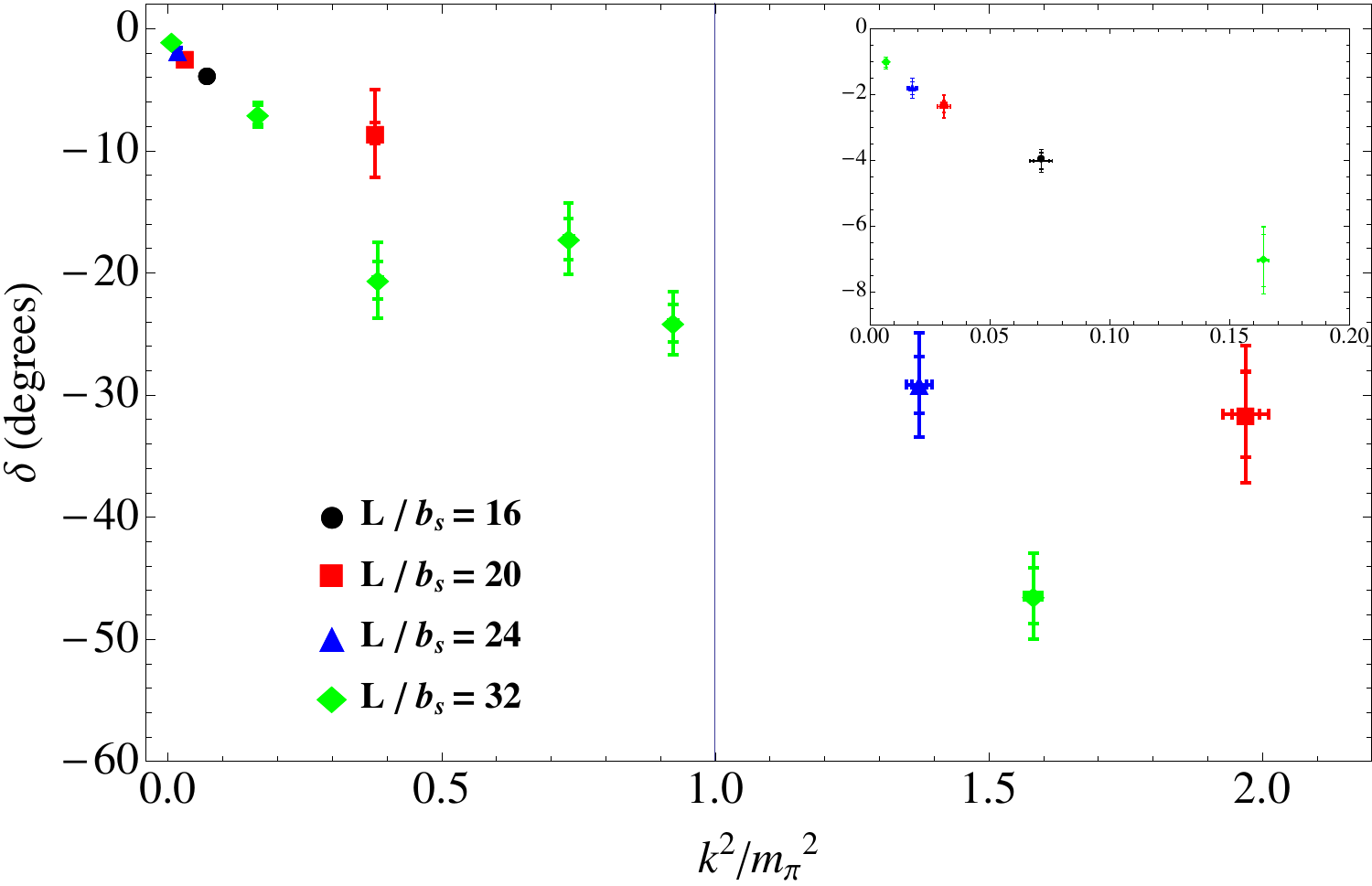}
\caption{Results of the Lattice QCD calculations
processed through the energy-eigenvalue relation to give values of 
the phase-shift $\delta$. The phase-shift at low energies is shown as an inset. 
The \{circles, squares, triangles, diamonds\}
(\{black, red, blue, green\}) correspond to 
the ensembles 
\{$16^3\times 128$, $20^3\times 128$, $24^3\times 128$, $32^3\times 256$\}. 
Statistical and systematic uncertainties are shown as the inner and outer
error-bars, respectively.
The vertical (blue) line at $k^2=m_\pi^2$ indicates the limit of the 
range of validity of the ERE  set by the t-channel cut.
The inelastic threshold is at $k^2=3 m_\pi^2$.
}
\label{fig:phasedata}
\end{figure}

Note that while the $32^3\times 256$ $P_{cm}=\sqrt{2},n=0$ and
$20^3\times 128$ $P_{cm}=0,n=1$ levels appear discrepant, we believe
this is likely a statistical fluctuation.  Also, the phase
shift we have extracted from the first excited state in the
$24^3\times 128$ ensemble disagrees with the equivalent extraction
presented in Ref.~\cite{Dudek:2010ew}.  While we find a phase shift of
$\delta=-29.2\pm 2.3\pm 4.3^o$ at $k^2\sim 0.21~{\rm GeV}^2$,
Ref.~\cite{Dudek:2010ew} finds $\delta\sim -13\pm 2^o$ at $k^2\sim
0.2~{\rm GeV}^2$. Our result is consistent with the phase shifts at
the nearby momenta calculated on the $32^3\times 256$ ensemble.

\subsection{The Effective Range Expansion Parameters}

\noindent 
The ERE is an expansion of the real part of the inverse scattering amplitude in powers of
the CoM energy,
\begin{equation}
\frac{k\,\cot{\delta}}{m_\pi} \ =\  -\frac{1}{m_\pi a} 
\ +\ \frac{1}{2}\;m_\pi r\;\left(\frac{k^2}{m_\pi^2}\right) \ +\ 
P\; (m_\pi r)^3\;\left(\frac{k^2}{m_\pi^2}\right)^2 +\ldots
\label{eq:effrange}
\end{equation}
where $m_\pi a$ and $m_\pi r$ are the scattering length and effective
range in units of $m_\pi^{-1}$, and $P$ is the shape parameter. Here
$k=|{\bf k}|$ is the magnitude of each pion's momentum in the CoM. 
Such an expansion is expected to be convergent for energies below the t-channel
cut, which is set by $\pi\pi$ exchange in the t-channel.
The t-channel cut starts at ${k^2}={m_\pi^2}$, 
while the inelastic threshold is ${k^2}=3{m_\pi^2}$.

\begin{figure}[!ht]
  \centering
     \includegraphics[width=0.50\textwidth]{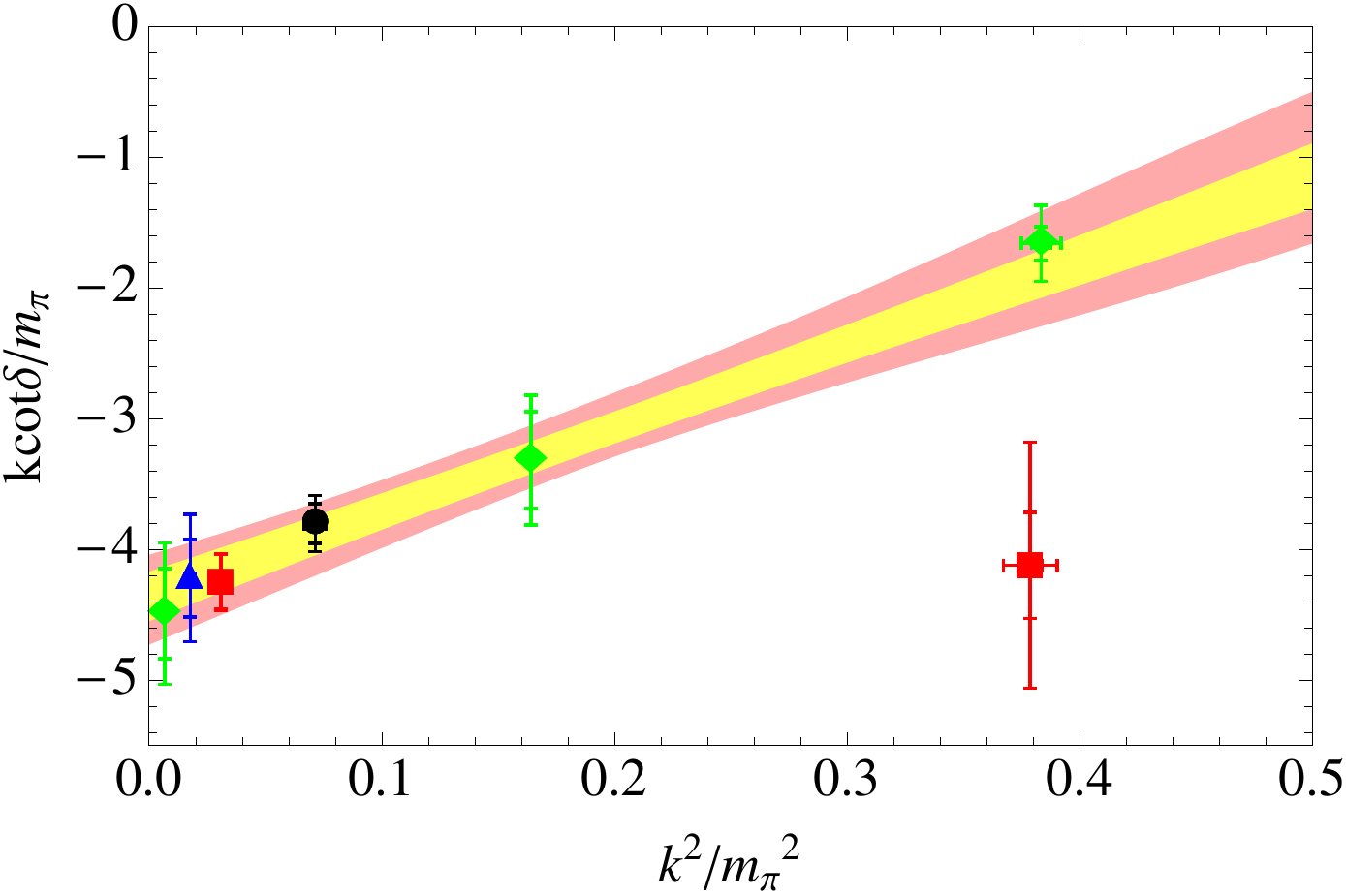}\ \ \
     \includegraphics[width=0.44\textwidth]{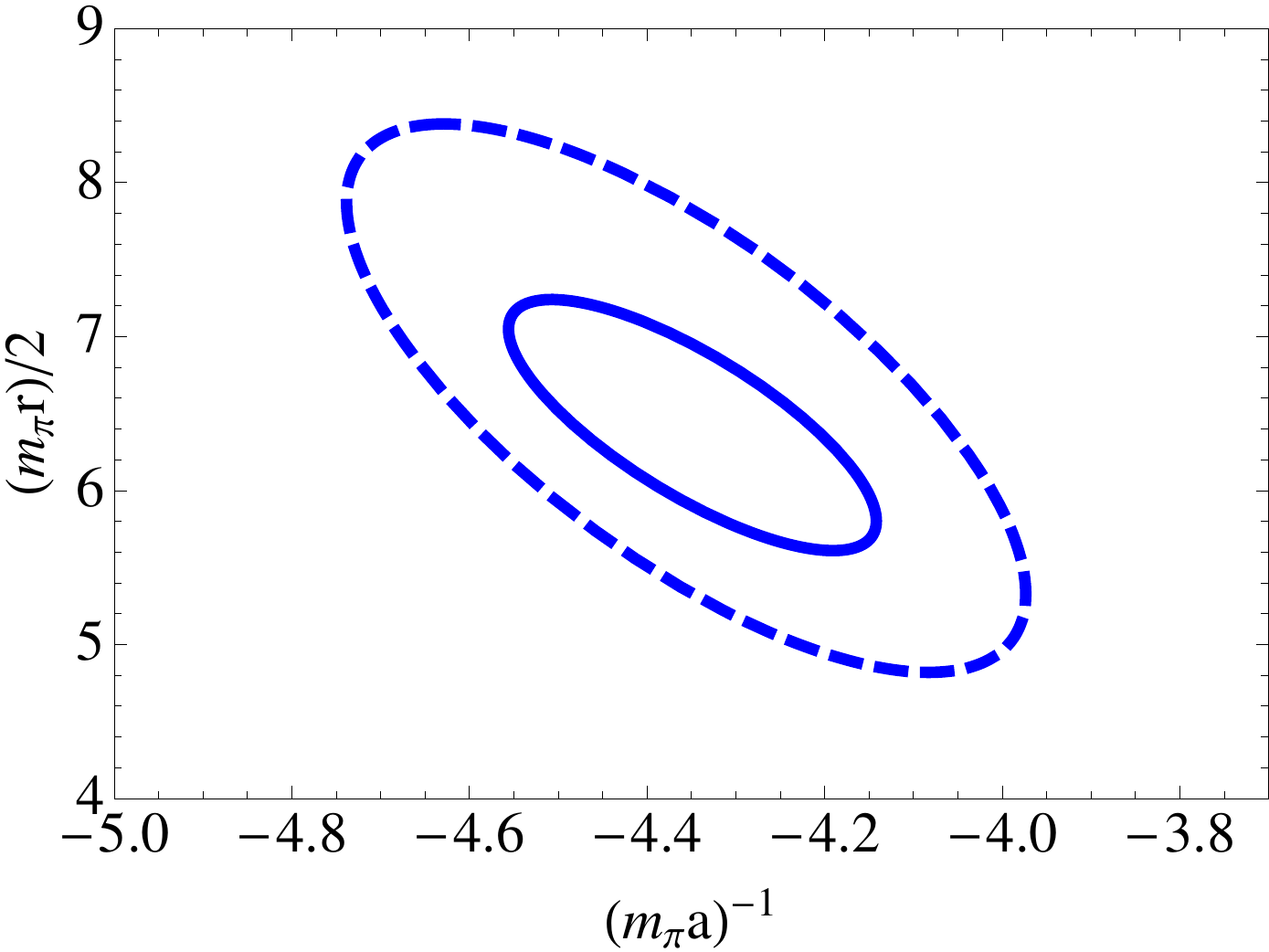}
     \caption{Left panel: a two-parameter fit to $k\cot\delta/m_\pi$
       over the region $k^2/m_\pi^2<0.5$ (Fit A).  The \{circles,
       squares, triangles, diamonds\} (\{black, red, blue, green\})
       correspond to the \{$16^3\times 128$, $20^3\times
       128$, $24^3\times 128$, $32^3\times 256$\} ensembles.  The shaded bands
       correspond to statistical (inner-yellow) and statistical and
       systematic added in quadrature (outer-pink).  Right panel:
       $68\%$ confidence interval error ellipses in the $(m_\pi
       a)^{-1}$ - ${1\over 2} m_\pi r$ space.  The inner-solid ellipse
       and the outer-dashed ellipse correspond to the statistical and to the
       statistical and systematic uncertainties added in quadrature, respectively.}
\label{fig:pipiEREfitA}
\end{figure}

\begin{figure}[!ht]
  \centering
     \includegraphics[width=0.50\textwidth]{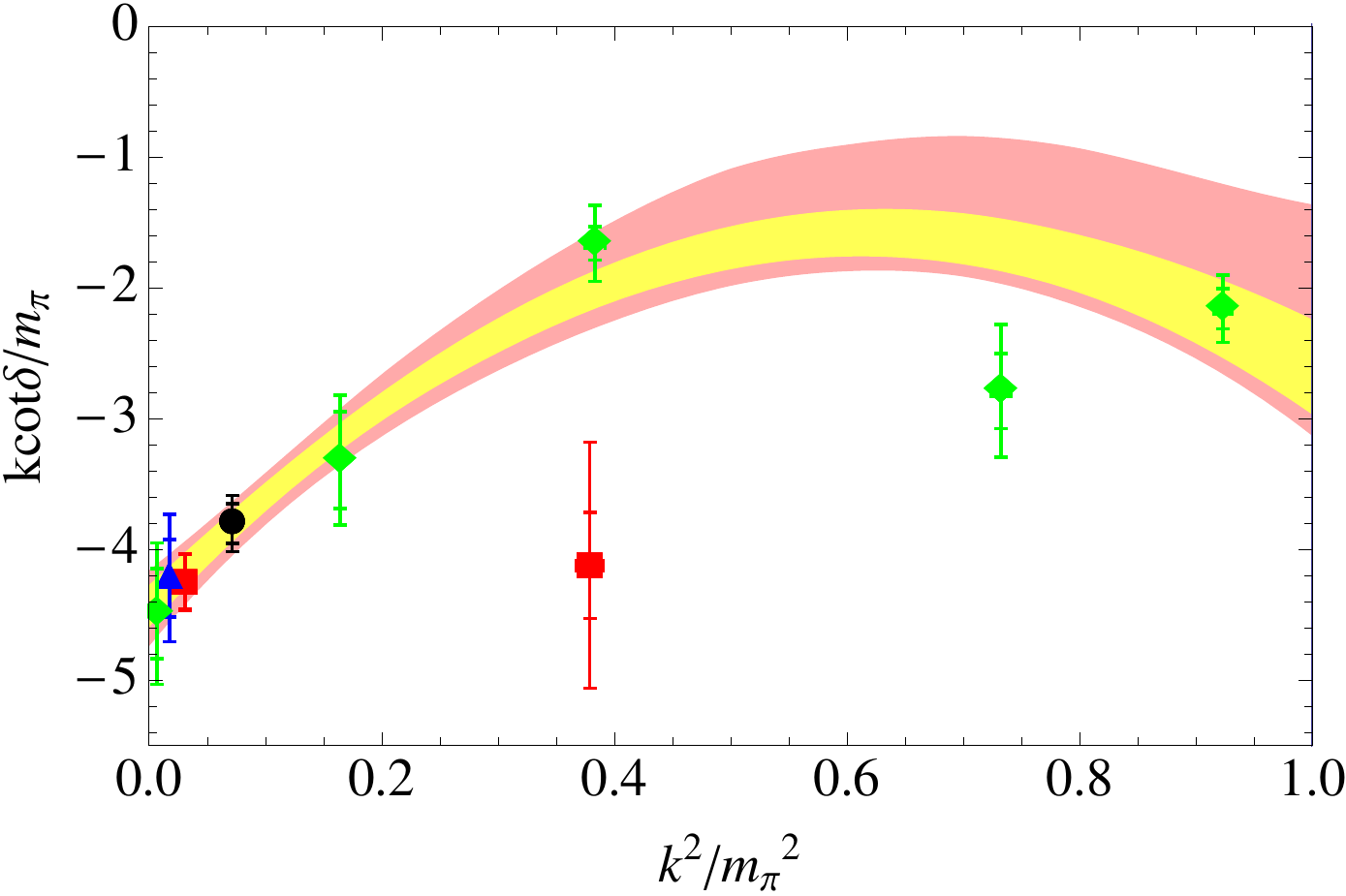}\ \ \
     \includegraphics[width=0.44\textwidth]{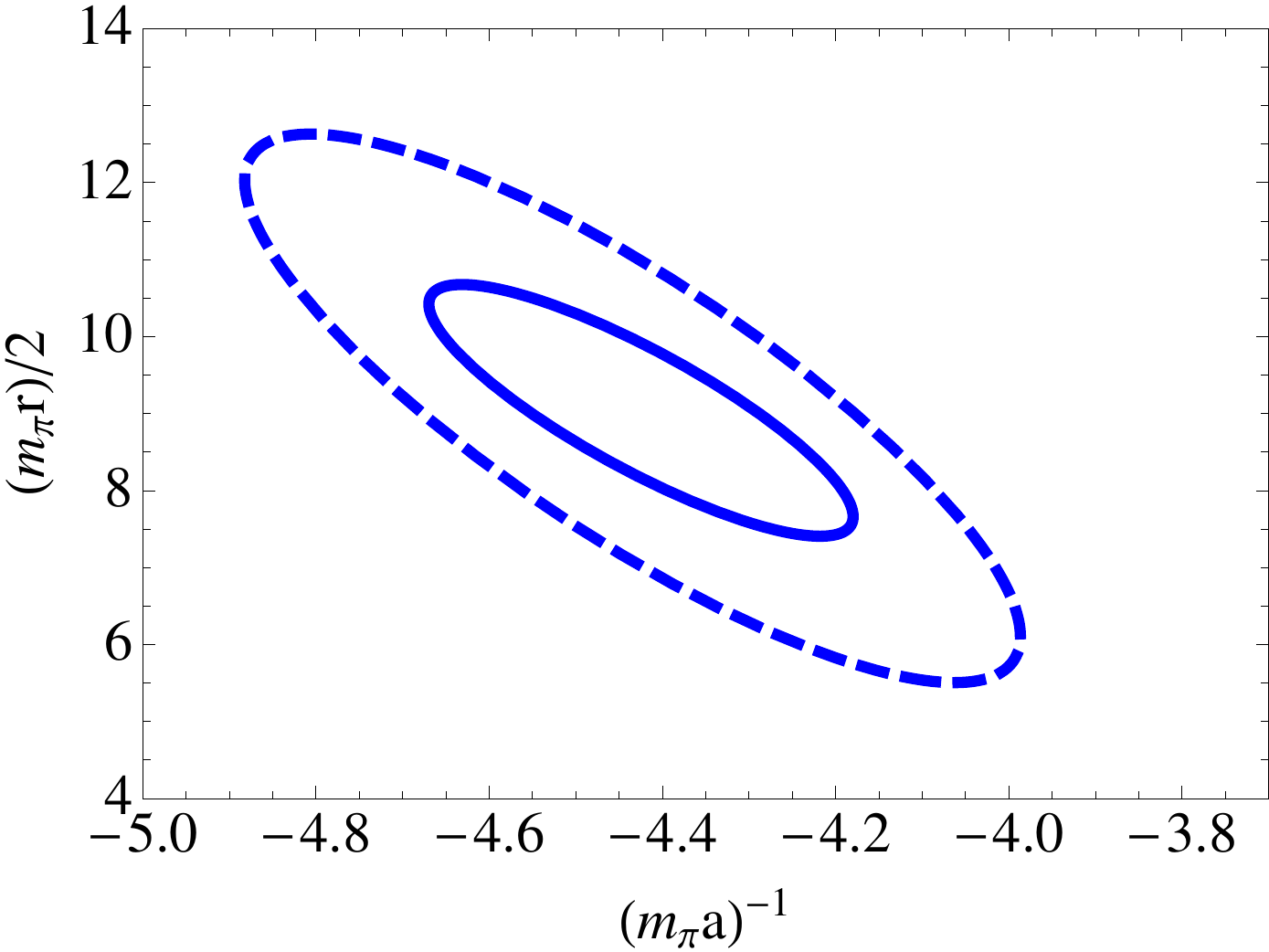}
     \caption{Left panel: a three-parameter fit to $k\cot\delta/m_\pi$
       over the region $k^2/m_\pi^2<1$ (Fit B).  The \{circles,
       squares, triangles, diamonds\} (\{black, red, blue, green\})
       correspond to the \{$16^3\times 128$, $20^3\times
       128$, $24^3\times 128$, $32^3\times 256$\} ensembles.  The shaded bands
       correspond to statistical (inner-yellow) and statistical and
       systematic added in quadrature (outer-pink).  Right panel:
       $68\%$ confidence interval error ellipses in the $(m_\pi
       a)^{-1}$ - ${1\over 2} m_\pi r$ space.  The inner-solid ellipse
       and the outer-dashed ellipse correspond to the statistical and to the
       statistical and systematic uncertainties added in quadrature, respectively.}
\label{fig:pipiEREfitB}
\end{figure}

As the calculations of $k\cot\delta/m_\pi$ are approximately linear in $k^2$
in the region $k^2/m_\pi^2<0.5$, the scattering length and
the effective range are fit (Fit A) using eq.~(\ref{eq:effrange}) with $P$ and
the other higher order terms set to zero.  The extracted values of
$m_\pi a$ and $m_\pi r$ are given in Table~\ref{tab:fitstoERT}, and
the resulting fit is shown in fig.~\ref{fig:pipiEREfitA}, along with
the $68\%$ confidence interval error ellipses for the two-parameters.
In the region $k^2/m_\pi^2<1$ the Lattice QCD calculations exhibit
curvature consistent with quadratic (and higher) dependence on $k^2$.
In Fit B the three leading ERE parameters are fit to the results of
the Lattice QCD calculations.  The fits are compared to the Lattice
QCD calculations in fig.~\ref{fig:pipiEREfitB}, which also shows the
$68\%$ confidence interval error ellipse for the two-parameter
subspace of the three-parameter fit.  It is clear from
Table~\ref{tab:fitstoERT} that the fit parameters are consistent
within the combined statistical and systematic uncertainties.  In what
follows, where we use $\chi$PT to predict the parameters at the
physical point, the spread in value of the ERE parameters will serve
as a useful gauge of the systematic uncertainty introduced in the
fitting of the scattering amplitude.  It is noteworthy that the data
allows a significant determination of the shape parameter, $P$.
\begin{table}[ht]
\caption{ERE parameters extracted from the Lattice QCD
  calculations of  $k\cot\delta/m_\pi$.}
\label{tab:fitstoERT}
\begin{tabular}{@{}|c | c | c | c | }
\hline
\quad Quantity \quad  & \quad Fit A: $k^2/m_\pi^2<0.5$ \quad & \quad Fit B: $k^2/m_\pi^2<1$ \quad  \\
\hline
$m_\pi a$      & 0.230(10)(16)   &  0.226(10)(16) \\
\hline
$m_\pi r$      & 12.9(1.5)(2.9)     & 18.1(2.4)(4.7)  \\
\hline
$m^2_\pi a r$  & 2.95(20)(42)     &  4.06(30)(57)   \\ 
\hline
$P$            & ---     &  -0.00123(30)(55) \\
\hline
\hline
$\chi^2/{\rm dof}$ & 0.83     &  0.79 \\
\hline
\end{tabular}
\end{table}

\section{Chiral  Interpolations}
\label{sec:chiextrap}

\subsection{Motivation}

\noindent 
Although these Lattice QCD calculations have been performed only at
one value of the pion mass, as we will see, the effective range and
threshold scattering parameters satisfy low-energy theorems mandated
by chiral symmetry, and therefore each scattering parameter can be
used to fix the corresponding LEC that appears at NLO in $\chi$PT.
Thus the scattering parameters at the physical point can be predicted
at NLO in $\chi$PT.  This is, in a sense, a chiral interpolation
rather than an extrapolation since one is interpolating between the
pion mass of the Lattice QCD calculation and the chiral limit.
Unfortunately, the pion decay constant, $f_\pi$, has not yet been
accurately computed on the anisotropic lattice ensembles that have
been used in this work.  However, $\chi$PT and the results of
mixed-action Lattice QCD calculations~\cite{NPLQCDforth} can be used
to determine $f_\pi$ (and its uncertainty) evaluated at the pion mass
of the present Lattice QCD calculations up to lattice spacing artifacts.
Specifically, in what follows we use $\sqrt{z^{\rm latt}}\equiv
m_\pi^{\rm latt}/f_\pi^{\rm latt}=2.59(13)$ at $m_\pi\sim 390~{\rm
  MeV}$.

\subsection{Threshold Parameters in $\chi$PT}
\noindent 
The relation between the $\pi^+\pi^+$ s-wave scattering amplitude $t(s)$
($= t_{L=0}^{I=2}(s)$)  
and the phase shift $\delta$
is given by~\cite{Bijnens:1997vq}
\begin{eqnarray}
t(s)=\left(\frac{s}{s-4}\right)^{1/2}\frac{1}{2i}
\{e^{2i\delta(s)}-1\}\ ,
\label{eq:unitary}
\end{eqnarray}
where $s= 4 (1 + k^2/m_\pi^2)$ and $k=|{\bf k}|$ is the magnitude of  the three-momentum
of each $\pi^+$ in the CoM frame.
The NLO scattering amplitude can be expressed in terms of three LEC's,  $C_1$, $C_2$, and 
$C_4$~\cite{Gasser:1983yg,Bijnens:1995yn}:
\begin{eqnarray} 
\label{eq:pipinlo}
t(k) 
& = & -\frac{m_\pi^2}{8\pi f_\pi^2}
-\frac{m_\pi^4}{f_\pi^4}\left(\, C_1\ -\ \frac{31}{384\pi^3}\,\right) \nn 
&&\ +\ \frac{k^2}{f_\pi^2}\bigg\lbrack\, -\frac{1}{4\pi}
+\frac{m_\pi^2}{f_\pi^2} 
\left(\, \frac{301}{1152\pi^3}\ -\ \frac{1}{128\pi^2} C_2 -\frac{7}{2} C_1\,\right) \,\bigg\rbrack \nn
&&\ +\ \frac{k^4}{f_\pi^4}\bigg\lbrack\, \frac{14}{45\pi^3}\ -\ 
\left(\, \frac{19}{8} C_1\ -\ \frac{9}{512\pi^2} C_2 \ +\  216\pi C_4\,\right) \,\bigg\rbrack \nn
&&\ -\ \frac{1}{4\pi^3 f_\pi^4} \left(\, \frac{3}{32} m_\pi^4\ 
+\ \frac{5}{12}m_\pi^2 k^2 \ +\ \frac{5}{9} k^4\,\right)\log\left(\frac{m_\pi^2}{f_\pi^2}\right) \nn
&&\ +\ \frac{1}{16 \pi^3 f_\pi^4} \left(\, \frac{1}{4} m_\pi^4\ 
+\ m_\pi^2 k^2 \ +\ k^4\,\right)
\sqrt{\frac{k^2}{k^2+m_\pi^2}}
\log\left(\frac{\sqrt{\frac{k^2}{k^2+m_\pi^2}}-1}{\sqrt{\frac{k^2}{k^2+m_\pi^2}}+1}\right) \nn
&&\ +\ \frac{1}{8\pi^3 f_\pi^4} \left(\, \frac{3}{16} m_\pi^4\ +\ \frac{7}{9}
  m_\pi^2 k^2 \ 
+\ \frac{11}{18}k^4\,\right)
\sqrt{\frac{k^2+m_\pi^2}{k^2}}
\log\left(\frac{\sqrt{\frac{k^2+m_\pi^2}{k^2}}-1}{\sqrt{\frac{k^2+m_\pi^2}{k^2}}+1}\right) \nn
&&\ -\ \frac{m_\pi^4}{128\pi^3 f_\pi^4} \left(\, 1 \ +\ \frac{13}{12}\frac{m_\pi^2}{k^2}\,\right)
\log^2\left(\frac{\sqrt{\frac{k^2+m_\pi^2}{k^2}}-1}{\sqrt{\frac{k^2+m_\pi^2}{k^2}}+1}\right)  \ .
\end{eqnarray}
The $C_i$ can be expressed in terms of the $l_i\equiv l_i^r(\mu=f_\pi)$, 
the familiar low-energy constants of two-flavor $\chi$PT~\cite{Gasser:1983yg},  
\bea 
C_1 &\equiv&-\frac{1}{2 \pi }(4\,{l_1}+4\,{l_2}+{l_3}-{l_4})-\frac{1}{128 \pi ^3} \ ; \nn
C_2 &\equiv& 32 \pi  (12\,{l_1}+4\,{l_2}+7\,{l_3}-3\,{l_4})+\frac{31}{6 \pi }\ ; \nn
C_4 &\equiv&\frac{1}{5184 \pi ^2}(212\,{l_1}+40\,{l_2}+123\,{l_3}-69\,{l_4})\ +
\ \frac{701}{622080 \pi^4}
\ .
\label{eq:Cs}
\eea

The behavior of the amplitude near threshold ($k^2\rightarrow 0$)
can be written as a power-series expansion in the CoM energy
\begin{eqnarray}
\mbox{Re}\; t(k)\ =\ -m_\pi a +k^2\;b + k^4\;c + O(k^6),
\label{eq:threshexp} 
\end{eqnarray}
where the threshold parameters $b$ and $c$ are referred to as slope parameters. 
Matching the threshold expansion in eq.~(\ref{eq:threshexp}) to the ERE  in
eq.~(\ref{eq:effrange}) gives~\cite{Bijnens:1997vq}:
\begin{equation}
m_\pi r \ =\ -\frac{1}{m_\pi a}-\frac{2 m^2_\pi b}{(m_\pi a)^2}+2 m_\pi a \ ;
\end{equation}
\begin{equation}
P = -\frac{(m_\pi a)^3\big\lbrack (m_\pi a)^2-4 (m_\pi a)^4+8(m_\pi a)^6-4\left(m_\pi a+2(m_\pi a)^3\right)b\;m_\pi^2-8\left(b^2+m_\pi a\; c\right)m_\pi^4\big\rbrack}
{8(m_\pi a-2(m_\pi a)^3+2b\; m_\pi^2)^3}  .
\end{equation}
These equations can be inverted to obtain $b$ and $c$ 
from the lattice-determined ERE  parameters.
Expanding the NLO amplitude in eq.~(\ref{eq:pipinlo}) in powers of $k$,
one finds NLO $\chi$PT expressions for the ERE and threshold parameters:
\bea
m_\pi a  &=& \frac{z}{8\pi}\ +\ z^2\,C_1\ +\ \frac{3z^2}{128\pi^3}\log z 
\ \ ,\ \ 
m_\pi r \ = \  \frac{24\pi}{z}\ +\ C_2\ +\ \frac{17}{6\pi}\log z \ , \nn
m_\pi^2 a r &=& 3\ +\ z\,C_3\ +\ \frac{11z}{12\pi^2}\log z
\ \ ,\ \ 
P \ = \  -\frac{23z^2}{13824\pi^2}\ +\ z^3\,C_4\ +\ \frac{613z^3}{995328\pi^4}\log z \ , \nn
m_\pi^2 b  &=& -\frac{z}{4\pi}\ -\ z^2\,\left(\,\frac{7}{2}\,C_1 \ +\ \frac{1}{128\pi^2}\,C_2 \ +\ 
\frac{5}{48\pi^3}\log z   \right) \ , \nn
m_\pi^4 c  &=& -z^2\,\left(\,\frac{19}{8}\,C_1 \ -\ \frac{9}{512\pi^2}\,C_2 \ +\ 216\pi\,C_4 \ +\
\frac{5}{36\pi^3}\log z   \right) \ ,
\label{eq:extrapforms}
\eea
where $z\equiv m_\pi^2/f_\pi^2$ and $C_3 =  24\pi C_1 + {1\over 8\pi}C_2 $.
It is important to note that the shape parameter $P$ and the threshold parameter $c$
do not receive contributions from LO $\chi$PT; i.e. they vanish in current algebra.

\subsection{Chiral Interpolation of Threshold Parameters}

\begin{figure}[!ht]
  \centering
     \includegraphics[width=0.90\textwidth]{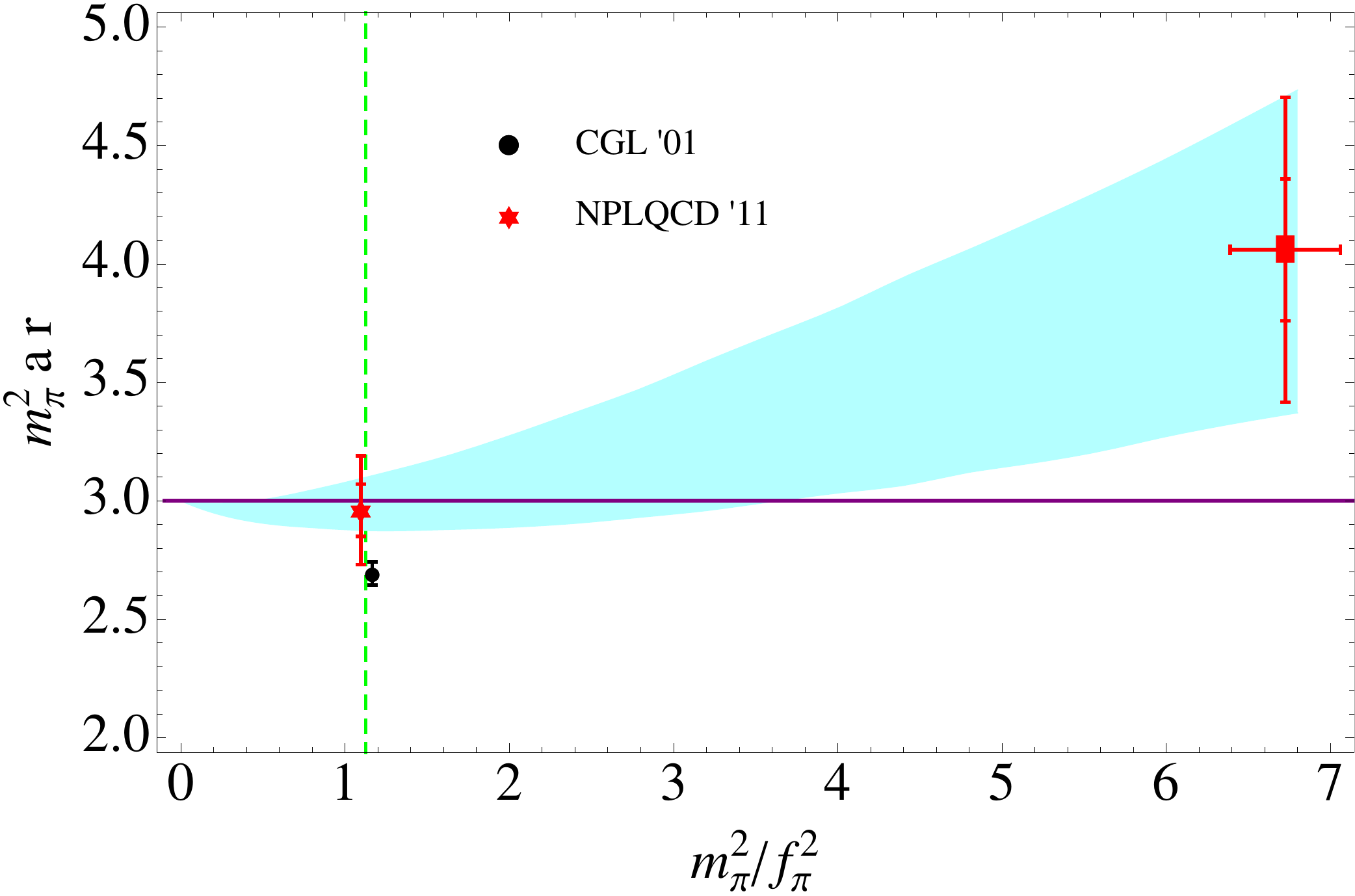}
\caption{The dashed (green) line denotes the physical line, and the horizontal
solid (purple) line denotes the LO $\chi$PT prediction, which is $m_\pi^2 a r=3$ in the
chiral limit. The band denotes the $68\%$ confidence interval interpolation
of the results of the lattice calculation (the (red) rectangle) from Fit B. 
The Lattice QCD $+\chi$PT prediction at the physical point is the (red) star on the
physical line, and the Roy equation prediction~\protect\cite{Colangelo:2001df}
is the (black) circle on the physical line.}
\label{fig:mamrPHYS}
\end{figure}

\noindent 
Using the ERE parameter set from Fit B given in table~\ref{tab:fitstoERT}, 
with statistical and
systematic uncertainties  combined  in quadrature, the four
functions $C_1((m_\pi a)^{latt},z^{latt})$, $C_2((m_\pi
r)^{latt},z^{latt})$, $C_3((m_\pi^2 a r)^{latt},z^{latt})$,
$C_4(P^{latt},z^{latt})$ can be determined. 
The ERE parameters  in table~\ref{tab:fitstoERT} give
\begin{eqnarray}
C_1^{NLO} & = & -0.00237(52) 
\ \ ,\ \ 
C_2^{NLO} \ =\  5.2(5.2) \; ;\nn
C_3^{NLO} & = & -0.02(0.10) 
\ \ ,\ \ 
C_4^{NLO} \ =\  9.0(4.0)\times 10^{-6} 
\ , 
\label{eq:Csfit}
\end{eqnarray}
where the superscripts denote that these constants are evaluated at NLO
in $\chi$PT, and from which follow, using eq.~(\ref{eq:extrapforms}), the predictions at the
physical point\footnote{Note that the precise NPLQCD result for the scattering length, $m_\pi
a=-0.04330(42)$, computed in Ref.~\cite{Beane:2007xs} with domain-wall valence quarks on
staggered sea quarks, is more precise than the result of eq.~(\ref{eq:erepfits}).} of
\begin{eqnarray}
m_\pi a & =& 0.0417(07)(02)(16)
\ \ ,\ \ 
m_\pi r \ = \  72.0(5.3)(5.3)(2.7)
\ \ ,\ \ 
m_\pi^2 a r \ = \ 2.96(11)(17)(11)
\ , 
\nn
P & =& -2.022(58)(12)(76)\times 10^{-4} \ ,
\nn
b & = & -0.832(50)(0)(31)\times 10^{-1 }\ m_\pi^{-2}
\ \ ,\ \ 
c  \ = \ 0.013(33)(01)(0)\ m_\pi^{-4}
\ ,  
\label{eq:erepfits}
\end{eqnarray}
where the first systematic uncertainty has been estimated by comparing
the interpolated results of Fits A and B, and by ``pruning'' the
highest energy datum from the Lattice QCD results and refitting.  The
second systematic uncertainty contains estimates, using naive
dimensional analysis (NDA), of the effects from higher orders in the chiral
expansion, next-to-next-to-leading order (NNLO) and higher, as well as
the contributions from lattice spacing artifacts that are expected to
contribute at ${\cal O}(b_s^2)$~\cite{Buchoff:2008ve}. 
The chiral interpolation of $m_\pi^2 a r$ is shown in
fig.~\ref{fig:mamrPHYS}.  (Note that the band in
fig.~\ref{fig:mamrPHYS} represents Fit B, and the outer uncertainty on
the interpolated result represents the effect of the two systematic
uncertainties and the statistical uncertainty added in quadrature, as
described above.)

With eq.~(\ref{eq:Cs}), the fit values of the $C_i$ in
eq.~(\ref{eq:Csfit}) can be used to constrain various combinations of
the $l_i$, and the renormalization group can be used to express these
constraints in terms of the scale-independent dimensionless barred
quantities, the $\overline{l}_i$~\cite{Gasser:1983yg}~\footnote{ The
  $l_i$ are related to the $\overline{l}_i$ via
\begin{eqnarray}
l_i & = & {\gamma_i\over 32\pi^2}
\left( \overline{l}_i +  \log\left({m_\pi^2\over\mu^2}\right)
\ \right)
\ \ , \nonumber
\end{eqnarray}
where $\gamma_1 = {1\over 3}$, $\gamma_2 = {2\over 3}$, 
$\gamma_3 = -{1\over 2}$ and $\gamma_4 = 2$.
}. 
We find that
\begin{eqnarray}
\overline{l}_3-4\overline{l}_4 & =&  -29(27)  
\ \ ,\ \  
\overline{l}_1-6\overline{l}_4 \  = \  -32(25)  \nn 
2\overline{l}_1-3\overline{l}_3 & =&  28(29)  
\ \ ,\ \ 
\overline{l}_1+4\overline{l}_2 \  = \  15.8(6.7)  \ ,
\end{eqnarray}
where statistical and systematic uncertainties have been combined  in quadrature.
With increased precision in the determination of the ERE
parameters, such  determinations of the LECs could become competitive
with other methods.

These results may seem surprisingly accurate for a Lattice QCD
calculation performed at a single pion mass.  As mentioned previously,
it is the chiral symmetry constraints on the scattering parameters in
the approach to the chiral limit that is responsible for the
precision.  The scattering length obtained here
is consistent within uncertainties with the previous Lattice QCD
determinations~\cite{Beane:2005rj,Beane:2007xs,Feng:2009ij}.  Further,
the scattering length and threshold parameters are found to agree with
determinations from the Roy equation (with chiral symmetry
input)~\cite{Colangelo:2001df},
\begin{eqnarray}
m_\pi a  &=& 0.0444(10) 
\ \  ,\ \ 
b \ = \  -0.803(12) \times 10^{-1}m_\pi^{-2}\ ; \nn
m_\pi^2 a r  &=& \ 2.666(0.083)
\ ,
\end{eqnarray}
at the $1\sigma$-level. Fig.~\ref{fig:mamrPHYS} provides a comparison
of the lattice calculation (and interpolation) and the Roy equation
value of $m_\pi^2 a r$.

\subsection{Chiral Interpolation of the Phase Shift}
\noindent 
The $\pi^+\pi^+$ scattering phase-shifts calculated with Lattice QCD,
which extend above the range of validity of the ERE but remain below
the inelastic threshold, can be used to predict the phase shift at the
physical value of the pion mass.  While the chiral expansion may break
down for scattering at sufficiently high energies, we ignore this
issue and fit the NLO $\chi$PT amplitude (one-loop level) to the
results of the Lattice QCD calculations at all of the calculated
energies, the maximum invariant mass being $\sqrt{s} \sim 1340~{\rm
  MeV}$.

\begin{figure}[!ht]
  \centering
     \includegraphics[width=0.49\textwidth]{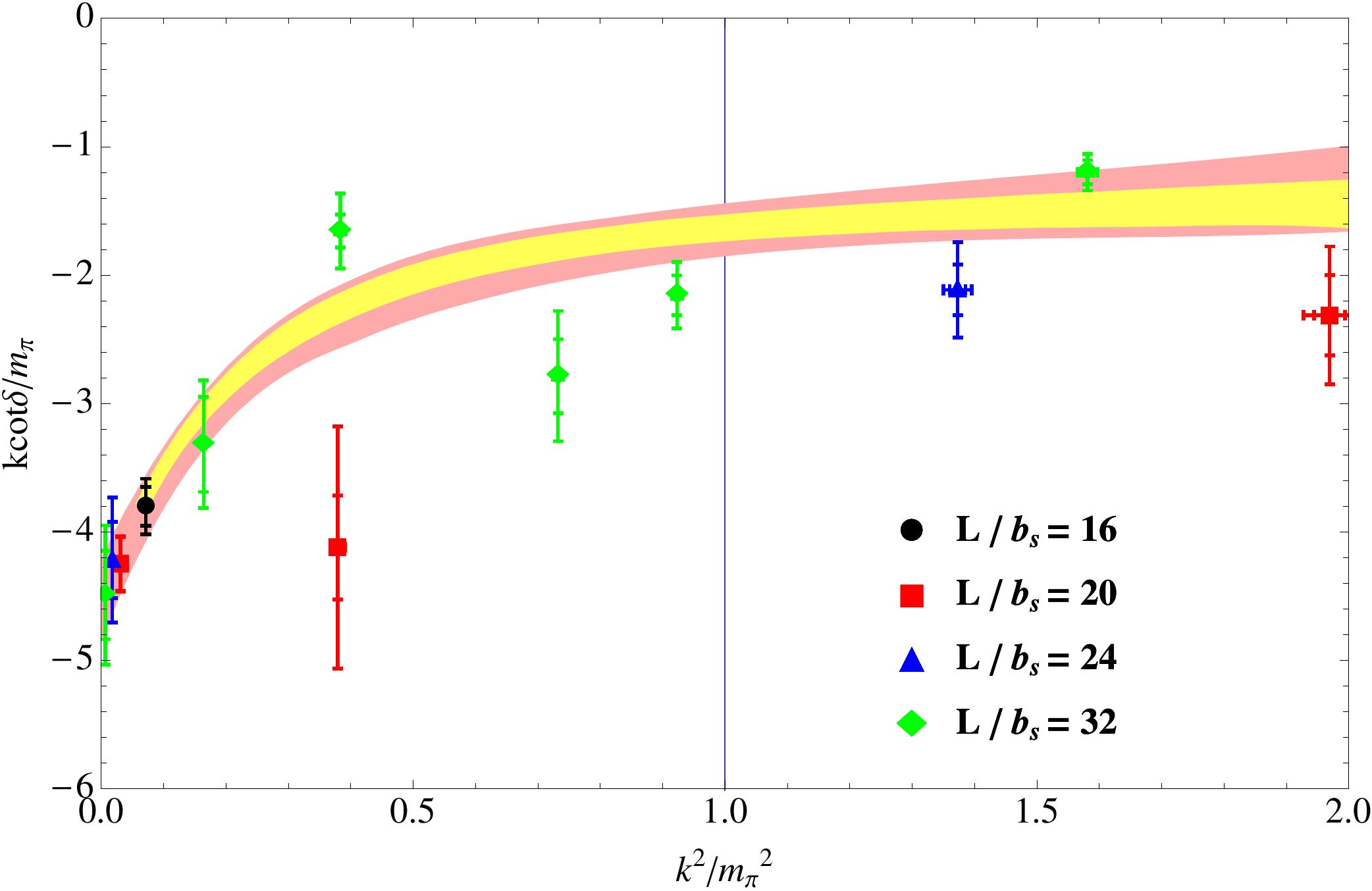}
     \includegraphics[width=0.49\textwidth]{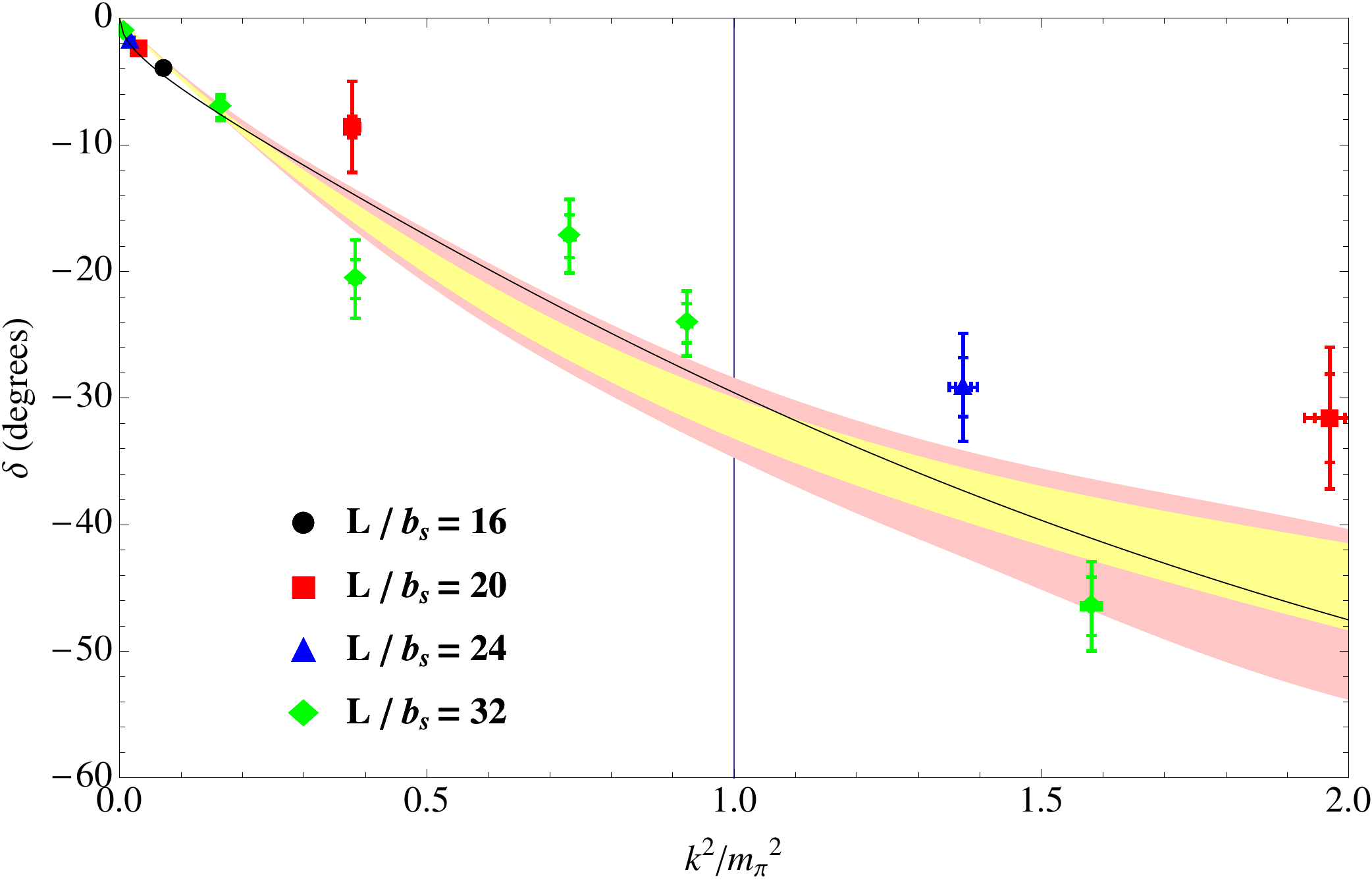}
\caption{Three parameter fit ($C_1$, $C_2$, and $C_4$)
of the NLO $\chi$PT expression for $k\cot\delta/m_\pi$
to the results of the Lattice QCD calculations.
The shaded bands correspond to statistical (inner-yellow) uncertainties and
statistical and systematic uncertainties added in quadrature (outer-pink). 
The solid (black) curve in the right panel is the LO $\chi$PT prediction (current algebra) 
at the pion mass used in the Lattice QCD calculations, $m_\pi \sim 390~{\rm MeV}$.}
\label{fig:pipiphasefit}
\end{figure}

\begin{figure}[!ht]
  \centering
     \includegraphics[width=0.90\textwidth]{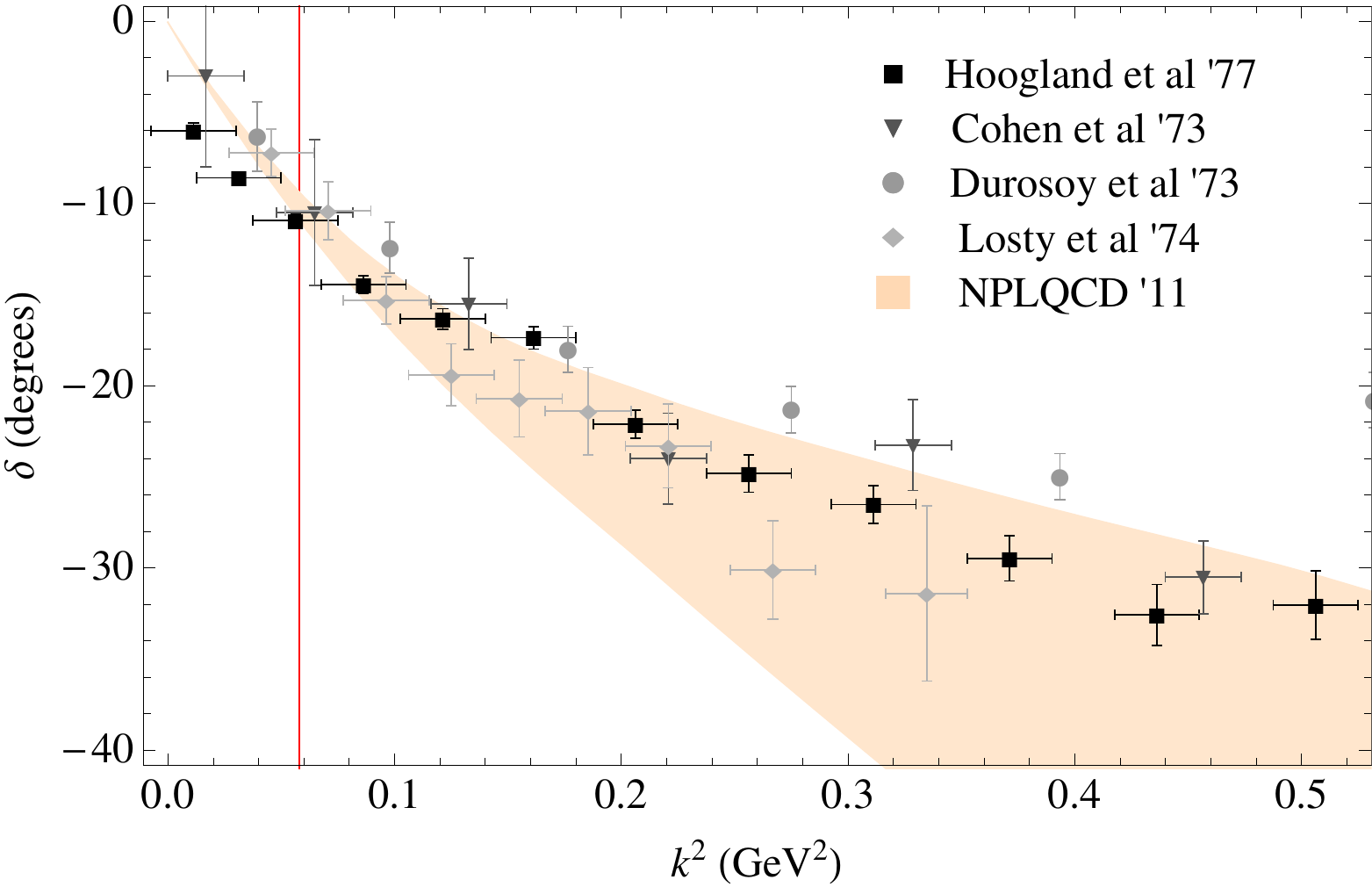}
\caption{The shaded band is the 
Lattice QCD  prediction of the  phase shift at the physical value of
  the pion mass, $m_\pi\sim 140~{\rm MeV}$ using  NLO $\chi$PT with the 
statistical and systematic uncertainties combined  in quadrature. 
The data is experimental (black and grey) taken from Refs.~\protect\cite{Hoogland:1977kt, Cohen:1973yx, Durusoy:1973aj,Losty:1973et}.
The red vertical line denotes the inelastic ($4\pi$) threshold.}
\label{fig:pipiphasephys}
\end{figure}

\begin{figure}[!ht]
  \centering
     \includegraphics[width=0.90\textwidth]{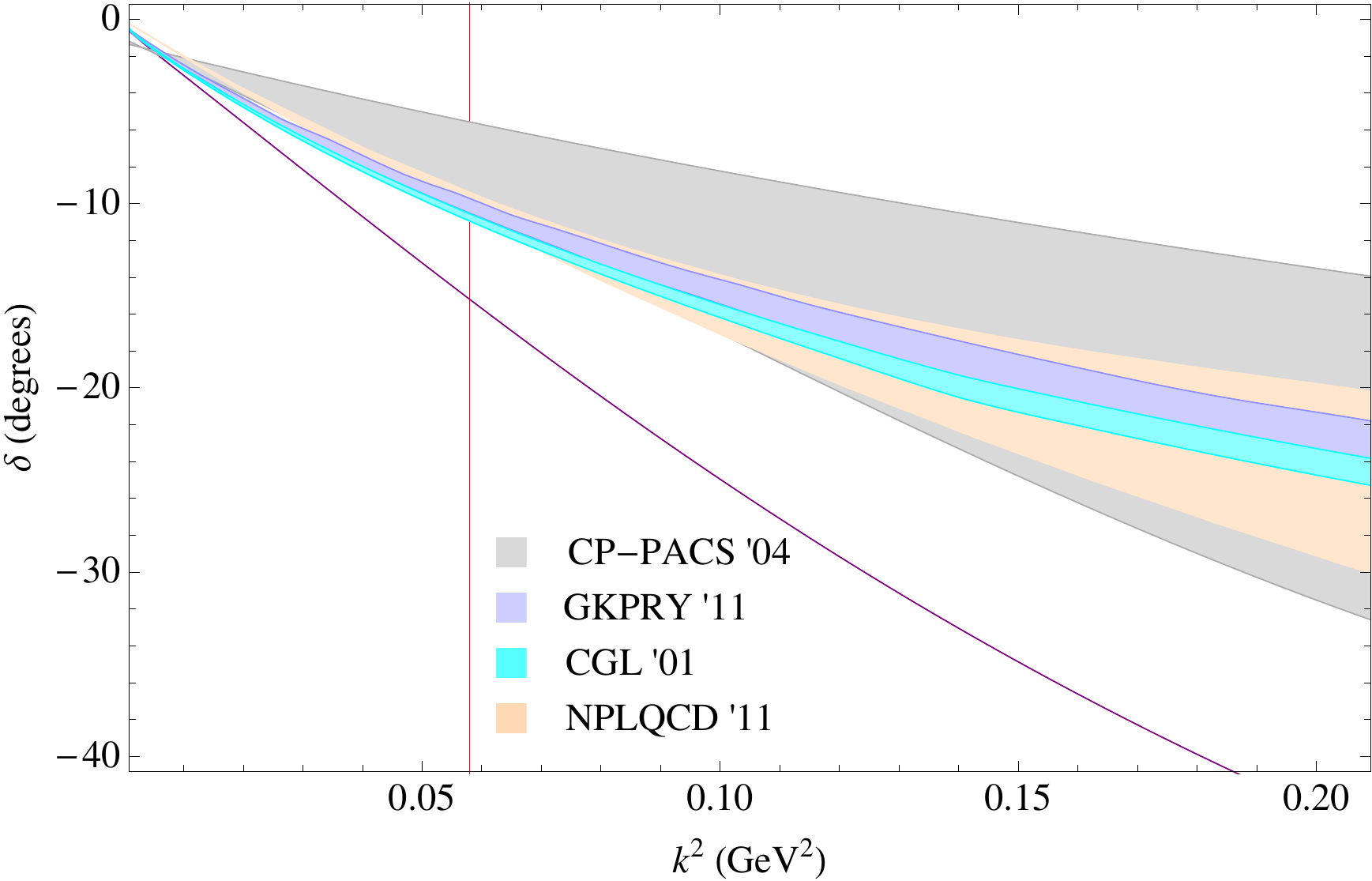}
     \caption{The shaded band is the Lattice QCD prediction of the
       phase shift at the physical value of the pion mass, $m_\pi\sim
       140~{\rm MeV}$ using NLO $\chi$PT with the statistical and
       systematic uncertainties combined in quadrature.  The outer
       (magenta) band is the CP-PACS physical
       prediction~\protect\cite{Yamazaki:2004qb}, and the two inner
       (blue and purple) bands are the Roy equation
       predictions~\protect\cite{Colangelo:2001df,GarciaMartin:2011cn}.
       (Note that the band from Ref.~\cite{Colangelo:2001df} lies above
       the band from Ref.~\cite{GarciaMartin:2011cn}.)
       The solid (purple) curve is the LO $\chi$PT prediction (current
       algebra) at the physical pion mass. The red vertical line
       denotes the inelastic ($4\pi$) threshold.}
\label{fig:pipiphasephysTH}
\end{figure}

The results of the Lattice QCD calculations given in
Table~\ref{tab:LQCDpionpion} 
are fit to the formula
\begin{equation}
\frac{k\,\cot{\delta}}{m_\pi} \ =\  \sqrt{1+\frac{k^2}{m_\pi^2}}\;
\left(\ \frac{1}{t_{\rm LO}(k)} \ -\ \frac{t_{\rm NLO}(k)}{
(t_{\rm LO}(k))^2} \right) \ +\ i\;\frac{k}{m_\pi} \ ,
\label{eq:chiptfitform}
\end{equation}
where $t_{\rm LO}$ and $t_{\rm NLO}$ are the LO and NLO contributions
to $t(k)$ in the chiral expansion, given in eq.~(\ref{eq:pipinlo}).
The result of the fit is shown in fig.~\ref{fig:pipiphasefit}; in the
left panel the fit (of $C_1$, $C_2$, and $C_4$) to $k\cot\delta/m_\pi$
is shown, and in the right panel, the fit values of $C_1$, $C_2$, and
$C_4$ (fully correlated) are used to predict the phase shift at the
pion mass of the Lattice QCD calculations, $m_\pi \sim 390~{\rm MeV}$.
The $68\%$ confidence intervals for $C_1$, $C_2$, and $C_4$ from this
fit are
\begin{eqnarray}
C_1^{NLO} &=& \left( -0.0040 , -0.0013 \right)
\ \ ,\ \
C_2^{NLO}\ \ =\ \ \left( 2.67 , 24.1 \right)  \ ,\ \nonumber \\
C_4^{NLO} & = &   \left( -1.7  , +3.6 \right)\times 10^{-5} 
 \ ,
\end{eqnarray}
with a $\chi^2/{\rm dof} = 2.1$ (for the fit with the statistical and
systematic uncertainties combined in quadrature).  The interpolated
ERE parameters are:
\begin{eqnarray}
m_\pi a \ =\ 0.0412(08)(16)
\ \ ,\ \ 
m_\pi r \ = \  80.0(9.58)(3.0)
\ \ ,\ \ 
P\  =\  -1.85(31)(07)\times 10^{-4} \ ,
\end{eqnarray}
which are consistent within uncertanties, but less precise, than the
threshold determinations of eq.~(\ref{eq:erepfits}).  Here the
second uncertainty is an NDA estimate of the effects of higher orders
in the chiral expansion and lattice spacing artifacts.  For a better
determination of the threshold parameters from the global fit, one
requires more accurate Lattice QCD calculations and the $\pi^+\pi^+$
amplitude beyond NLO in the chiral expansion. In
fig.~\ref{fig:pipiphasephys} the fit values of $C_1$, $C_2$, and $C_4$
are used to predict the phase shift at the physical value of the pion
mass, $m_\pi\sim 140~{\rm MeV}$, which is compared to the experimental
data of Refs.~\cite{Hoogland:1977kt, Cohen:1973yx,
  Durusoy:1973aj,Losty:1973et}.  Fig.~\ref{fig:pipiphasephysTH}
compares the phase shift prediction to the Lattice QCD phase-shift
determination by CP-PACS~\cite{Yamazaki:2004qb}, and the Roy equation
determinations of the phase shift from
Refs.~\cite{Colangelo:2001df,GarciaMartin:2011cn}.  One should keep in
mind that the interpolated phase shift is valid above the inelastic
threshold, as the $4\pi$ intermediate state appears beyond NLO in the
$\chi$PT calculation (at two-loop level).  The combined Lattice QCD
and $\chi$PT prediction of the $\pi^+\pi^+$ phase shift at the
physical pion mass is found to be in good agreement with the
experimentally-determined phase shift.  While for $|{\bf k}| \gsim
400~{\rm MeV}$ the uncertainty in the prediction exceeds the
uncertainties in the experimental data, below this momentum the
Lattice QCD+$\chi$PT prediction is more precise.

It is interesting to observe that while the LO phase-shift well
reproduces the results of the Lattice QCD calculations at $m_\pi\sim
390~{\rm MeV}$, as shown in fig.~\ref{fig:pipiphasefit}, the NLO
contributions are important at the physical pion mass, as seen in
fig.~\ref{fig:pipiphasephysTH}.  This is to be contrasted with the
chiral behavior of the scattering length which is dominated by the LO
amplitude, with NLO making a small but noticeable contribution.  In an
attempt to isolate the origin of this apparent difference, it is
useful to consider scattering at NLO in the chiral limit where
\begin{eqnarray}
\cot\delta 
&\rightarrow &
-{4\pi f_\pi^2\over k^2}
\ +\ 
{40\over 9\pi}\log\left({2 k\over f_\pi}\right)
+38\pi^2 C_1 - {9\over 32} C_2 + 3456\pi^3 C_4 -{224\over 45\pi}
\ \ \ .
\end{eqnarray}
The phase-shift can be defined this way even in the chiral limit
because at LO and NLO the only intermediate states contributing to the
scattering amplitude involve two pions.  Inelastic channels, such as
four-pion intermediate states which would invalidate the relation in
eq.~(\ref{eq:unitary}), first contribute to the scattering amplitude
at NNLO.  This is what allows for the phase-shift to be predicted
above the inelastic threshold, and to remain perturbatively close
to the actual value for momenta below the chiral symmetry breaking
scale.  At LO in the expansion, the phase-shift reaches $\delta=\pi/4$
when $k^2=4\pi f_\pi^2\sim 0.22~{\rm GeV}^2$ (using $f_\pi=132~{\rm
  MeV}$), consistent with the phase-shift shown in
fig.~\ref{fig:pipiphasephysTH}.  Clearly, it is reasonable to take the
limit $m_\pi\ll k$ for this value of $k$ ($k\sim 470~{\rm MeV}$).
Further, at this value of $k$, the NLO terms are approximately
equal to the LO terms, providing an estimate of the 
convergence region of the chiral expansion for the scattering process.

It is also worth noting that while it is formally invalid to use the
L\"uscher relation in eq.(\ref{eqn:RumGott}) for the scattering of
pions above inelastic threshold, $\chi$PT indicates that the error
introduced into phase-shift determinations is small, occurring at NNLO
in the chiral expansion.  This is not expected to be true for other
scattering processes (those not involving the pseudo-Goldstone
bosons).  Therefore, while strictly speaking the results presented in
Ref.~\cite{Dudek:2010ew} above inelastic threshold arise from an
invalid application of eq.(\ref{eqn:RumGott}), the expected deviation
from the true result is expected to be small (at momenta for which the
chiral expansion is converging), suppressed by two orders in the
chiral expansion.  Clearly, precision calculations of the phase-shift
above the inelastic threshold cannot rely upon a methodology that does
not include the effects of inelastic processes.  As all of the
calculations in our work are below the inelastic threshold, the
present analyses and predictions do not suffer from this
inconsistency.

\section{Summary and Conclusion}
\label{sec:disc}

\noindent 
The increases in high-performance computing capabilities 
and the advent of powerful new algorithms have thrust
Lattice QCD into a new era where the interactions among hadrons can be
computed with controlled systematic uncertainties.  While 
calculation of the basic properties of nuclei and hypernuclei is now a
goal within reach, it is important to consider the simplest hadronic
scattering processes as a basic test of the lattice methodology for
extracting scattering information (including bound states) from the
eigenstates of the QCD Hamiltonian in a finite volume.  In this work,
we have calculated the $\pi^+\pi^+$ scattering amplitude using Lattice
QCD over a range of momenta below the inelastic threshold.  Our
predictions for the threshold scattering parameters, and hence the
leading three terms in the ERE expansion, are consistent with
determinations using the Roy equations~\cite{Colangelo:2001df,GarciaMartin:2011cn} and the
predictions of $\chi$PT.  In particular, our determination of $m_\pi^2
a r = 2.96(11)(17)(11)$ from an interpolation of a fit to the low momentum
values of $k\cot\delta/m_\pi$ is consistent with the LO prediction of
$\chi$PT of $m_\pi^2 a r = 3 \left( 1 + {\cal
    O}(m_\pi^2/\Lambda_\chi^2)\right)$.  Further, the resulting
predictions for the phase shift at the physical pion mass --using NLO
$\chi$PT-- are in agreement with experimental data, and are even more
precise in the low-momentum region. 

The Lattice QCD calculations presented here were performed at one
lattice spacing simply due to the lack of computational resources,
therefore, an extrapolation of the ERE parameters to the continuum
limit (as was performed in the work of CP-PACS~\cite{Yamazaki:2004qb})
could not be performed.  The discretization of the quark fields that
has been employed gives rise to lattice spacing artifacts at ${\cal
  O}(b_s^2)$, and we expect such contributions to be small for these
calculations.

\vfill\eject
\vspace{0.5in}
\noindent We thank G.~Colangelo, H.~Leutwyler, J.~Nebreda and
J.~Pelaez for valuable conversations and communications, K.~Roche for
computing resources at ORNL NCCS and R.~Edwards and B.~Joo for
developing QDP++, Chroma~\cite{Edwards:2004sx} and production.  We
acknowledge computational support from the USQCD SciDAC project, NERSC
(Office of Science of the DOE, Grant No.~DE-AC02-05CH11231), the UW
HYAK facility, Centro Nacional de Supercomputaci\'on (Barcelona,
Spain), LLNL, the Argonne Leadership Computing Facility at Argonne
National Laboratory (Office of Science of the DOE, under contract
No.~DE-AC02-06CH11357), and the NSF through Teragrid resources
provided by TACC and NICS under Grant No.~TG-MCA06N025.  SRB was
supported in part by the NSF CAREER Grant No.~PHY-0645570.  The Albert
Einstein Center for Fundamental Physics is supported by the
“Innovations- und Kooperationsprojekt C-13” of the “Schweizerische
Universit\"atskonferenz SUK/CRUS”.  The work of EC and AP is supported
by the contract FIS2008-01661 from MEC (Spain) and FEDER.  AP
acknowledges support from the RTN Flavianet MRTN-CT-2006-035482 (EU).
H-WL and MJS were supported in part by the DOE Grant
No.~DE-FG03-97ER4014.  WD and KO were supported in part by DOE Grants
No.~DE-AC05-06OR23177 (JSA) and No.~DE-FG02-04ER41302.  WD was also
supported by DOE OJI Grant No.~DE-SC0001784 and Jeffress Memorial
Trust, Grant No.~J-968.  KO was also supported in part by NSF Grant
No.~CCF-0728915 and DOE OJI Grant No.~DE-FG02-07ER41527.  AT was
supported by NSF Grant No.~PHY-0555234 and DOE Grant
No.~DE-FC02-06ER41443.  The work of TL was performed under the
auspices of the U.S.~Department of Energy by LLNL under Contract
No.~DE-AC52-07NA27344 and the UNEDF SciDAC Grant
No.~DE-FC02-07ER41457.  The work of AWL was supported in part by the
Director, Office of Energy Research, Office of High Energy and Nuclear
Physics, Divisions of Nuclear Physics, of the U.S. DOE under Contract
No.~DE-AC02-05CH11231.

\end{document}